\documentclass[10pt,reprint,amsmath,aps,pra]{revtex4-1}

\usepackage[latin1]{inputenc}
\usepackage[T1]{fontenc}
\usepackage{graphicx}
	\graphicspath{{./}{figures/}}
\usepackage{xcolor}		
\usepackage[colorlinks=true,linkcolor=purple,citecolor=blue,urlcolor=teal]{hyperref}

\usepackage{mathrsfs} 

\DeclareMathOperator{\sinc}{\mathrm{sinc}}
\DeclareMathOperator{\rect}{\mathrm{rect}}
\renewcommand{\j}{\mathrm{i}}
\renewcommand{\v}[1]{\mathbf{#1}}
\newcommand{\dirac}{\delta}
\renewcommand{\aa}{{\op{a}^{\phantom{\dag}}}}
\newcommand{\ac}{{\op{a}^{\dag}}}

\newcommand{\pol}{\vec{\boldsymbol{\varepsilon}}}
\newcommand{\rr}{\v{r}}
\newcommand{\kk}{\v{k}}

\newcommand{\aao}[1][]{{\op{o}_{#1}^{\phantom{\dag}}}}
\newcommand{\aco}[1][]{{\op{o}_{#1}^{\dag}}}

\newcommand{\vEp}{\v{E}^{(\! + \!)}}

\newcommand{\vopEp}{\op{\v{E}}^{(\! + \!)}}
\newcommand{\vopEm}{\op{\v{E}}^{(\! - \!)}}

\newcommand{\Ep}{E^{(\!+\!)}}

\newcommand{\TEp}{\breve{E}^{(\!+\!)}}

\newcommand{\chieff}{\chi_\text{eff}}

\newcommand{\rrho}{\boldsymbol{\rho}}
\newcommand{\kkappa}{\boldsymbol{\kappa}}

\newcommand{\gtemps}{T_p}
\newcommand{\gespace}{S_p}
\newcommand{\Tgtemps}{\breve{T}_p}
\newcommand{\Tgespace}{\breve{S}_p}
\newcommand{\energiepompe}{\mathscr{E}_p}
\newcommand{\fR}{\breve{R}}

\newcommand{\kpo}{k_{p_0}}

\newcommand{\opHam}{\op{\mathcal{H}}}

\newcommand{\tf}[1]{\breve{#1}}
\newcommand{\vol}{{V}}
\newcommand{\volq}{\mathcal{V}}

\newcommand{\intt}[1][]{\int_{#1}\,dt\:}
\newcommand{\intw}[1][]{\int_{#1}\:\frac{d\omega}{2\pi}\:}
\newcommand{\intr}[1][]{\int_{#1}\,d^3\rr\:}
\newcommand{\intk}[1][]{\int_{#1}\frac{d^3\kk}{(2\pi)^3}\:}

\newcommand{\nf}{n'}
\newcommand{\nfs}{n'_s}
\newcommand{\nfi}{n'_i}
\newcommand{\nc}{n}
\newcommand{\ns}{n_s}
\renewcommand{\ni}{n_i}
\newcommand{\np}{n_p}

\newcommand{\K}{K}
\newcommand{\Q}{Q}

\newcommand{\FF}{\mathcal{F}}

\renewcommand{\SS}{\mathcal{S}}
\newcommand{\LL}{\mathcal{L}}

\newcommand{\OO}{\mathcal{O}}
\newcommand{\TOO}{\breve{\OO}}

\newcommand{\vgp}{v_{p}}
\newcommand{\vgs}{v_{s}}

\newcommand{\vdelta}{\delta}
\newcommand{\vrho}{\alpha}
\newcommand{\veta}{\xi} 
\newcommand{\vzeta}{\zeta} 
\newcommand{\vphiz}{{\varphi_0}} 
\newcommand{\vvphis}{\boldsymbol{\varphi}_{s}}
\newcommand{\vvphii}{\boldsymbol{\varphi}_{i}}

\newcommand{\DwF}{\Delta \omega_\text{F}}
\newcommand{\Dwp}{\Delta \wp}
\newcommand{\Dtp}{\Delta t_p}

\newcommand{\w}{\omega}
\renewcommand{\wp}{\omega_p}
\newcommand{\wpo}{\omega_{p_0}}
\newcommand{\ws}{\omega_{s}}
\newcommand{\wso}{\omega_{s_0}}
\newcommand{\wi}{\omega_{i}}
\newcommand{\wio}{\omega_{i_0}}
\newcommand{\wF}{{\omega_\text{F}}}

\newcommand{\PAB}{P_\text{AB}}
\newcommand{\Int}[1][\omega]{\int\!\frac{d{#1}}{2\pi}\;}
\newcommand{\IInt}[1][\kkappa]{\int\!\!\!\int\!\!\frac{d^2{{#1}}}{(2\pi)^2}\;}

\newcommand{\ket}[1]{\ensuremath{| {#1} \rangle}}
\newcommand{\braket}[2]{\ensuremath{\langle #1 | #2 \rangle}}
\newcommand{\abs}[1]{\left| #1 \right|}
\newcommand{\op}[1]{\hat{#1}}
\newcommand{\conv}{\! * \!}

\newcommand{\Chi}{\,\overline{\overline{\boldsymbol{\chi}}}^{(2)}}

\newcommand{\thetas}{\theta_s}
\newcommand{\thetai}{\theta_i}
\newcommand{\rhos}{\rho_s}
\newcommand{\rhoi}{\rho_i}

\newcommand{\Cp}{C_p}

\begin{document}

\title{Optimal photon-pair single mode coupling in narrow-band spontaneous parametric down-conversion with arbitrary pump profile}

\author{Jean-Loup \surname{Smirr}}
\author{Matthieu \surname{Deconinck}}
\author{Robert \surname{Frey}}
\author{Imad \surname{Agha}}
\author{Eleni \surname{Diamanti}}
\author{Isabelle \surname{Zaquine}}
\email{isabelle.zaquine@telecom-paristech.fr}
\affiliation{Institut TELECOM/Telecom ParisTech -- CNRS/LTCI\\46 rue Barrault 75013 \textsc{Paris}, France}
\date{2011}

\begin{abstract}
	A theoretical study of the performance of single-mode coupled spontaneous parametric down-conversion sources is proposed, which only requires very few assumptions of practical interest : narrow-bandwidth and quasi-degenerate collinear generation. Other assumptions like pump beam spatial and temporal envelopes, target single-mode profile and size, and non-linear susceptibility distribution, are only taken into account in the final step of the computation, thus making the theory general and flexible. Figures of merit for performance include absolute collected brightness, pair collection efficiency and heralding ratio. The optimization of these values is investigated through functions that only depend on dimensionless parameters, allowing for deducing from the results the best experimental configuration for a whole range of design choices (e.g. crystal length, pump power). A particular application of the theory is validated by an experimental optimization obtained under compatible assumptions. A comparison with other works and proposals for numerically implementing the theory in its most generality are also provided.
\end{abstract}
	
\maketitle

\section{Introduction}\label{intro}

Sources of entangled photons have applications ranging from fundamental tests of quantum mechanics \cite{Tittel2001,Groblacher2007} to quantum information and communications \cite{Gisin2007,Gisin2010}. Entanglement based quantum key distribution protocols have been reported \cite{Ekert1991} as an alternative to those based on single photon sources and quantum repeaters have also been shown to require entanglement as a primary resource \cite{Shapiro2002,Simon2007}.

Spontaneous parametric down-conversion (SPDC) remains the simplest way to generate entangled photons, enabling telecom wavelength generation which cannot be easily achieved using atomic cascades. A large choice of configurations is available according to the emission geometry (colinear \cite{Wong2006} or not \cite{Kwiat1995,Noh2007}), the phase matching type (I \cite{Altepeter2005} or II \cite{Kwiat1995}) and the signal and idler frequencies (degenerate \cite{Shi2004} or not \cite{Guillet2006}). The non-colinear emission geometry has some advantages, providing direct separation of signal and idler photons but colinear emission in periodically poled crystals recently became increasingly popular \cite{Fiorentino2005,Fedrizzi2007,Kuzucu2008}, because of the high brightness obtained with longer crystals and higher nonlinear susceptibilities.

Filtering is one of the most important aspects of photon pair engineering. Spatial filtering and more specifically coupling the photon pairs to a single spatial mode, like an optical fiber, is the first requirement for long-distance communications. The coupling efficiency is then a key parameter, knowing that losses directly limit the maximum possible visibility of the source \cite{Virally2010,Smirr2010a}. Narrow-band spectral filtering can be necessary to ensure spectral indistinguishibility but also to obtain efficient coupling to a quantum memory \cite{Lvovsky2009,Simon2010,Chaneliere2010,Saglamyurek2011}. It is also beneficial for long-distance communications of polarization states as it reduces polarization mode dispersion.

Since the pioneering works of SPDC concerning the quantum fluctuation and noise in parametric processes \cite{Louisell1961} and those dealing with probability of coincidences in the emission of signal and idler photons \cite{Hong1985,Ghosh1987,Mandel1999} an important theoretical effort has been developed in order to optimize various kind of sources \cite{Rubin1994,Keller1997}, some of them dedicated to strongly focused beams \cite{Pittman1996}. 

After studies of the collection of SPDC through apertures \cite{Joobeur1994, Kurtsiefer2001}, numerous authors have reported studies of the coupling of SPDC into a single spatial mode like that of a fiber \cite{Bovino2003,Castelletto2004,Castelletto2005,Ljunggren2005,Ling2009,Mitchell2009,Bennink2010}, with various assumptions and experimental methods of validation.

The optimal focusing of a continuous wave monochromatic pump has been investigated \cite{Ljunggren2005} in order to calculate the maximum coupling efficiency of photon pairs into single-mode fibers but the pump diffraction was neglected and the phase mismatch was not taken into account in the optimization process. The same approximations have been used in \cite{Ling2009} to calculate the absolute emission rates of SPDC into single modes, where the crystal is moreover assumed to be thin. The absolute brightness is also calculated in \cite{Mitchell2009} but with a continuous wave pump, emphasizing the parallel between SPDC and the classical treatment used for second harmonic generation. 
Very recently, R. S. Bennink studied the case of Gaussian beam profiles, directly writing the nonlinear interaction hamiltonien with a gaussian beam profile for both the pump and the SPDC generated photon pairs \cite{Bennink2010}. In this case, neither the absolute brightness of the source nor the absolute collection efficiency of an optical fiber can be evaluated. Moreover, no spectral filtering is taken into account.

To our knowledge, a comprehensive study of narrow-band SPDC pumped by a diffracting beam of arbitrary pulse envelope and duration (from short pulses to continuous-wave lasers) and arbitrary transverse profile , covering a full range of focusing (from tight focusing to parallel beams) and arbitrary filter shapes, has not been addressed. Such a generalization can, however, be useful in view of optimizing future SPDC sources that might require specific spectral and spatial characteritics to interface with particular quantum information systems. 

In this paper, we propose a theoretical framework that is general enough to allow the investigation of these characteristics and their effects on the source performance (e.g. brightness, coupling efficiency) under assumptions suited to the desirable characteritics mentioned above : narrow-bandwidth for quantum information applications and colinear emission in long crystals with a given non-linear susceptibility distribution. Using a dedicated source, we also experimentally validate the theoretical predictions for particular assumptions.

Section \ref{sec:TheoreticalFramework} describes the general formalism of the addressed problem and derives the wave-function of the created photon pairs, taking into account spatial and spectral filtering. We also define various figures of merit that are useful to characterize a source performance. The numerical application of this general theoretical approach to the case of a narrow-band fibered source pumped by a Gaussian beam is developed in section \ref{sec:NarrowBand}. The experimental setup  is described in section \ref{sec:Experiment} and the measured performances are compared with the theoretical previsions. A comparison of our results with other works, as well as guidelines for generalizing the numerical calculations are proposed in section \ref{sec:Discussion}.

\section{Theoretical framework}\label{sec:TheoreticalFramework}

\begin{figure}
	\centering
	\includegraphics[width=\columnwidth]{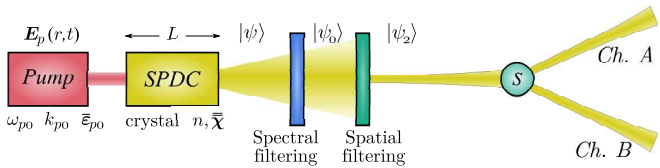}
	\caption{General configuration considered in the theoretical framework: a nonlinear crystal is pumped by a beam of arbitrary temporal and spatial profiles. The photon pairs produced through SPDC are spectrally filtered and coupled to a single spatial mode, before being split towards channels A and B.}
	\label{fig:IdealSetup}
\end{figure}
\subsection{The down-converted two-photon state}\label{par:DownConvertedState}

The general configuration considered hereafter is depicted in Figure \ref{fig:IdealSetup}. 
The non-linear interaction takes place in a crystal with a second-order susceptibility $\Chi(\rr)$, pumped by a classical field of positive-frequency complex amplitude $\vEp_p$. The  spontaneous parametric down-conversion process is described by a time-dependent Hamiltonian $\opHam(t)$. The signal and idler fields, ${\vopEp}_{s}$ and ${\vopEp}_{i}$, have the following plane-wave decomposition in the quantization volume $\volq$~:
\begin{equation}
	{\vopEp}_{s,i}(\rr,t) = \sum_{\ell_{s,i}} \pol_{\ell_{s,i}} \aa_{\ell_{s,i}} A(\ell_{s,i}) e^{\j (\kk_{\ell_{s,i}} \cdot \rr - \w_{\ell_{s,i}} t)}
\end{equation}
where the field amplitude $A(\ell) $ for the mode $\ell$ associated with annihilation operator $\aa_\ell$ is
\begin{equation}\label{eq:DefinitionA}
	A(\ell) = \j \sqrt{ \frac{ \hbar\w_{\ell}} {2 \epsilon_0 \volq n^2(\kk_{\ell})} },
\end{equation}

Frequency and polarization are respectively denoted by $\w_\ell$ and $\pol_\ell$, wavevectors in the crystal by $\kk_\ell$. A notation $\kk'_\ell$ will be used when the wavevector is evaluated in another medium.

In the interaction picture, the Hamiltonian is given by \cite{Boyd2008,Yariv1989} 
\begin{multline}\label{eq:InteractionHamiltionian}
	\opHam(t) = \epsilon_0 \intr \vEp_{p}(\rr,t) \, \Chi(\rr) \vopEm_s(\rr,t) \vopEm_i(\rr,t) \\ + \text{\emph{H.c.}}
\end{multline}
where $\vopEm$ is the Hermitian conjugate (\emph{H.c.}) of $\vopEp$. 

The pump field is assumed to be a linearly $\pol_x$-polarized classical paraxial beam propagating along the $z$ axis with $\mathrm{L}^2$-normalized temporal and spatial envelopes $\gtemps$ and $\gespace$.
 
Using transverse coordinates $\rrho = ( \begin{smallmatrix} x \\ y \end{smallmatrix} )$  so that $\rr = ( \begin{smallmatrix} \rrho \\ z \end{smallmatrix} )$
\begin{equation}\label{eq:fluo:ChampPompe}
	\vEp_p(\rrho,z,t) = \pol_x \Cp \gtemps \Big( t - \frac{z}{\vgp} \Big) \gespace(\rrho,z) e^{\j(\kpo z - \wpo t)}.
\end{equation}
Here $\kpo$ is the pump wavevector modulus at its central frequency $\wpo$ and $\vgp$ is the group velocity in the crystal. We neglect its spectral dispersion, therefore $\vgp = c/\np$ where $\np$ is the refractive index in the crystal at $\wpo$. The constant $\Cp = \sqrt{\energiepompe/(2\epsilon_0 \np^2 \vgp)}$ depends on the pump pulse energy $\energiepompe$.

When the down-conversion efficiency is small, the generated two-photon state is the first-order approximate solution to the Schrödinger equation. Discarding the zeroth-order term
\begin{equation}\label{eq:Schrodinger}
	\ket{\psi(t)} = \frac{1}{\j \hbar} \int^{t}_{-\infty} dt' \opHam(t') \ket{0}.
\end{equation}

Let us separate the spatial dependence $R(\rr)$ of the nonlinear susceptibility
\begin{equation}\label{eq:Chieff}
	\pol_{x} \Chi(\rr) \pol_{{\ell_s}} \pol_{{\ell_i}} = \chi(\ell_s,\ell_i) \cdot R(\rr).
\end{equation}
After the pump has propagated through the whole crystal length, ($t \rightarrow \infty$), the state is time-independent in the interaction picture. Identifying the integral in Eq. \eqref{eq:Schrodinger} as a Fourier transform of $\Ep_p = \vEp_p \cdot \pol_x$ with respect to its temporal variable and according to the Fourier transform conventions of Appendix \ref{ann:FourierConventions}, the state is given by
\begin{multline}\label{eq:LocalTwoPhotonState}
	\ket{\psi}= \frac{\epsilon_0}{\j \hbar} \sum_{\ell_s,\ell_i} \chi(\ell_s,\ell_i) A(\ell_s) A(\ell_i) \intr e^{-\j (\kk_{\ell_s}+\kk_{\ell_i}) \cdot \rr } \\
	\times R(\rr) \TEp_p (\rr,\w_{\ell_s}+\w_{\ell_i}) \ac_{\ell_s} \ac_{\ell_i} \ket{0}.
\end{multline}
This equation can be understood as the sum of the contributions of local interactions to the delocalized down-converted two-photon state $\ket{\psi}$. The integral has the form of a spatial Fourier transform, which leads to
\begin{equation}\label{eq:intLocalTwoPhotonState2}
	\ket{\psi} = \sum_{\ell_s,\ell_i} \gamma_0 (\ell_s,\ell_i) \ac_{\ell_s} \ac_{\ell_i} \ket{0},
\end{equation}
where
\begin{multline}\label{eq:gamma0}
	\gamma_0 (\ell_s,\ell_i)  = \frac{\epsilon_0}{\j \hbar} \chi(\ell_s,\ell_i) A(\ell_s) A(\ell_i) \\
		\times \fR \conv \TEp_p(\kk_{\ell_s}+\kk_{\ell_i},\w_{\ell_s}+\w_{\ell_i}),
\end{multline}
$\fR$ being the Fourier transform of $R$. Using the fact that, for a diffracting beam in the paraxial approximation, the Fourier transform of the transverse spatial envelope has the following $z$ dependence
\begin{equation}\label{eq:Diffracting beam}
	\Tgespace(\kkappa,z) = \Tgespace(\kkappa,0) e^{- \j \frac{|\kkappa|^2}{2 \kpo} z},
\end{equation}
where $\kkappa$ is the transverse wavevector, we have, according to the Fourier transform conventions of Appendix \ref{ann:FourierConventions}~:
\begin{multline}
	\TEp_p(\kkappa,k_z,\w) = \Cp \Tgtemps(\w - \wpo) \Tgespace(\kkappa,0) \\
		 \qquad \times 2\pi \dirac \big( k_z - \kpo - \frac{\w - \wpo}{\vgp} + \frac{\abs{\kkappa}^2}{2 \kpo}\big).
\end{multline}

We consider a crystal of length $L$ centered on $z=0$ whose transverse dimensions are large compared to the pump beam profile. If its nonlinear susceptiblity distribution is transversally invariant, it can be expressed as a one-dimensional Fourier series
\begin{equation}
R(z) = \rect(z/L) \times \sum_{m} R_m e^{-\j 2 \pi m \frac{z}{L}},
\end{equation}
where $\rect(z)$ is the rectangular function, with $\rect(0)=1$. In the reciprocal space, 
\begin{equation}
\fR(k_z) = \sum_{m} R_m \sinc \left( k_z + 2 \pi m/L \right).
\end{equation}
Then $\gamma_0(\ell_s,\ell_i) = \sum_m {\gamma_0}_m (\ell_s,\ell_i)$ with
\begin{multline}\label{eq:fluo:gamma0}
	{\gamma_0}_m (\ell_s,\ell_i)  = \frac{\epsilon_0}{\j \hbar} R_m \chi(\ell_s,\ell_i) A(\ell_s) A(\ell_i) \\
	\times \Cp \Tgtemps(\w_{\ell_s}+\w_{\ell_i}-\wpo) \Tgespace(\kkappa_{\ell_s}+\kkappa_{\ell_i},0) \\
	\times L \sinc \left( \Delta K_m(\kk_{\ell_s}+\kk_{\ell_i},\w_{\ell_s}+\w_{\ell_i}) \frac{L}{2} \right),
\end{multline}
where 
\begin{equation}\label{eq:fluo:DeltaK}
	\Delta K_m(\kk,\w) = k_z-\kpo - \frac{\w-\wpo}{\vgp} + \frac{\abs{\kkappa}^2}{2 \kpo} + m \frac{2 \pi}{L}.
\end{equation}

So far, we have explicited in Eq.\eqref{eq:intLocalTwoPhotonState2} and \eqref{eq:gamma0} the unnormalized state $\ket{\psi} $, of a photon-pair down-converted during a pump pulse. Its squared modulus $\braket{\psi}{\psi}$ is the probability for such a down-conversion to effectively occur. The effect of spectral and spatial filtering on the photon-pair state is developped in the following section.

\subsection{Spectral and spatial filtering}\label{par:Filtering}

The calculation of the effect of filtering on one-photon states is detailed in Appendix \ref{ann:OpticalElements}. Extrapolation to two-photon states is straightforward and only results are given here. Filters give rise to losses and two photon states are consequently transformed into mixed states, i.e. non coherent superpositions of two-photon, one-photon and zero-photon terms. In the case of coincidence counting, the two-photon state component is post-selected and we will discard zero- and one-photon terms.

Purely spectral filters are described by a function $\FF(\w-\wF)$ defining their amplitude transmission with a maximum normalized at $1$ at their central frequency $\wF$. More generally, a spectral filter can be sensitive to the direction of the wavevector (like prisms, gratings or Fabry-Pérot etalons), and it may be necessary to take a $\kkappa$-dependence into account~: $\FF(\w-\wF, \kkappa)$.

Spatial filtering is modelled as the coupling of the down-converted field to a single spatial mode defined, in a medium of refractive index $\nf$, by a frequency dependent function $\OO_{\w,0}(\rr)$. The filter location $z=z_0$ defines the transverse plane where the coupling is considered to take place. This coupling induces a state projection on the target mode, and the state of the transmitted photons is associated with a creation operator $\aco[\w]$.

Following Eq.~\eqref{eq:intLocalTwoPhotonState2}, the spectrally-filtered, free-space two-photon state is given by
\begin{equation}\label{eq:SpectrallyFilteredState}
	\ket{\psi_0} = \sum_{\ell_s,\ell_i} \gamma_0 (\ell_s,\ell_i)\gamma_T^{(2)}(\ell_s,\ell_i) \ac_{\ell_s} \ac_{\ell_i} \ket{0},
\end{equation}
where
\begin{equation}\label{eq:gammaT}
	\gamma_T^{(2)}(\ell_s,\ell_i) = \FF(\w_{\ell_s}-\wso,\kkappa_{\ell_s}) \FF(\w_{\ell_i}-\wio,\kkappa_{\ell_i}),
\end{equation}
and $\wso$ and $\wio$ are the signal and idler frequencies selected by the filters, which can be equal for degenerate down-conversion using a unique filter. The squared modulus of this state, $\braket{\psi_0}{\psi_0}$, is the probability $P_0$ of generating a photon pair during a pump pulse in the filter bandwidth.

The spectrally and spatially filtered state is given by
\begin{equation}\label{eq:CoupledState}
	\!\!\ket{\psi_2} = \sum_{\ell_s,\ell_i} \gamma_0 (\ell_s,\ell_i)\gamma_T^{(2)} (\ell_s,\ell_i)\gamma_S^{(2)}(\ell_s,\ell_i) \aco[\w_{\ell_s}] \aco[\w_{\ell_i}] \ket{0},
\end{equation}
where
\begin{multline}\label{eq:gammaS}
	\gamma_S^{(2)}(\ell_s,\ell_i) = \frac{1}{\SS}\TOO^*_{\w_{\ell_s},0}(\kkappa_{\ell_s},z_0) \TOO^*_{\w_{\ell_i},0}(\kkappa_{\ell_i},z_0) \\
\times e^{\j (k'_{z,\ell_s} + k'_{z,\ell_i}) z_0},
\end{multline}
$\TOO^*_{\w,0}$ is the complex conjugate of the transverse Fourier transform of $\OO_{\w,0}$,  $k'_{z,\ell} = k_{z,\ell} \nf(\w_\ell)/\nc(\w_\ell)$ is the longitudinal wavevector evaluated in medium of refractive index $\nf$, and $\SS$ is the transverse section of the quantization volume.

Due to spectral and spatial filtering, $\ket{\psi}$ has become $\ket{\psi_2}$. The squared modulus of this state, $\braket{\psi_2}{\psi_2}$, is the probability $P_2$ that one pulse has generated a photon pair and that this pair has been transmitted by the spectral and spatial filters. Its numerical calculation is performed in subsection \ref{Fom}, using assumptions that will allow us to separate the frequency dependence from the wavevector dependence in Eq. \eqref{eq:fluo:DeltaK} as shown in the next subsection.

\subsection{Assumptions}

A few assumptions are quite natural when aiming at applications in quantum information. Moreover, they enable further analytical development as well as faster numerical calculation. The consequent approximations, mostly used to simplify the expressions of $\Delta K$ in \eqref{eq:gammaS}, aim at eventually decoupling the frequency dependence ($\w$) from the angular dependence ($\kkappa$) in the wavevectors $\kk$.

\subsubsection{Collinear collection}

When using long crystals, as in most recent and efficient devices \cite{Fiorentino2004,Hentschel2009}, a collinear configuration is required. In such a case, the paraxial approximation, as already used for the pump beam,  can be applied to the down-converted photons on the $z$-axis, giving the following expression for the longitudinal component of the signal wavevector

\begin{equation}\label{eq:kz}
	{k_z}_s = k_{s_0} + \frac{\ws - \wso}{\vgs} - \frac{1}{2} \frac{ |\kkappa_s|^2 }{|\kk_s|},
\end{equation}
where $k_{s_0}$ is the wavenumber of collinearly emitted signal photons at the central frequency of the filter $\wso$, and $\vgs = c/\ns$ with $\ns$ the refractive index in the crystal at the signal frequency. A similar expression applies for ${k_z}_i$. 

\subsubsection{Narrow bandwidth}

When filtering limits the source bandwidth to less than a few nanometers ($\DwF \ll \wpo$), we have $\ws \approx \wso$ and $\wi \approx \wio$ in which case, according to Eq. \eqref{eq:kz}, the phase mismatch $\Delta K$ depends only on $\kkappa_s, \kkappa_i$. The fields amplitudes and nonlinear susceptibility, respectively defined in Eqs. \eqref{eq:DefinitionA} and \eqref{eq:Chieff}, can then be considered constant. 

A narrow bandwidth is a desirable characteristic for long-distance communications, as it reduces dispersion in optical fibers and is also required for projects of repeaters based on quantum memories \cite{Gisin2007,Simon2010}. In such memories based on atomic and ionic resonances, the acceptance bandwidth is lower than a few GHz \cite{Saglamyurek2011}, which definitely lies within the present assumption.

\subsubsection{Spatial dependance of the nonlinear susceptibility}
Although it is possible to calculate $\sum_m {\gamma_0}_m$ for an arbitrary $R(z)$, this function is generally chosen so that one term ${\gamma_0}_{\tilde{m}}$ is as high as possible and is the only optimally phase-matched one. This is achieved by periodically poling the nonlinear susceptibility with a period $\Lambda$. In this case, $\tilde{m}$ should be chosen such that $L = \tilde{m} \Lambda$, and $R_{\tilde{m}} = 2/\pi$, is the first term of the Fourier expansion of a square periodic function. In the following, we will consider the case of a collinear interaction in a periodically poled crystal with an effective susceptibility $\chieff=\chi(\wso,\wio) 2/\pi$, corresponding to specific polarizations of the pump, signal and idler beams. The case of the homogeneous crystal could also be described using an infinite period and $R_{\tilde{m}} = 1$.

\subsubsection{Quasi-degenerate down-conversion}

In the following, we will restrict the process to quasi-degenerate down-conversion, that is $| \delta \w | \ll \wpo$ with $\delta \w = \wso-\wio$. 

The phase mismatch then reduces to
\begin{multline}\label{eq:DeltaKcoldeg}
	\Delta K(\kkappa_s,\kkappa_i)\approx \Delta {k_0} + \frac{|\kkappa_s + \kkappa_i|^2}{2 \kpo} \\
		- \Big( 1 - \frac{\delta \w}{\wpo} \Big) \frac{\np}{\ns} \frac{|\kkappa_s|^2}{\kpo}
		- \Big( 1 + \frac{\delta \w}{\wpo} \Big) \frac{\np}{\ni} \frac{|\kkappa_i|^2}{\kpo},
\end{multline}

where $\Delta {k_0} = k_{s_0} + k_{i_0} - \kpo + \frac{2 \pi}{\Lambda}$ is the longitudinal phase mismatch. This together with the narrow bandwidth assumption allows for omitting the frequency dependence of the functions $\OO_{\w,0}$, describing the target mode. It can therefore be replaced by the unique function $\OO_0 = \OO_{\wpo/2,0} \approx \OO_{\wso,0} \approx \OO_{\wio,0}$.

\subsection{Figures of merit}\label{Fom}

In our general configuration of Fig. \ref{fig:IdealSetup}, only the spectrally and spatially filtered pairs with a probability $P_2$ will give rise to measured coincidences between the two channels. It is also important to be aware of the total brightness, accounting for all down-converted pairs in the spectral bandwidth under consideration $P_0$. The pair coupling efficiency is defined as the ratio $\Gamma_2 = P_2/P_0$. This energetical figure of merit, however, cannot describe the loss of coherence due to spatial filtering, related to the ratio of coincidences to single counts. That is why we are interested in the probability $\Gamma_{2|1}$ to have the idler photon coupled to the target spatial mode when the signal is coupled to that mode (or vice versa). It is sometimes called heralding ratio or conditional efficiency, measuring the ability of one photon to herald its twin for single photon source applications. It is also very useful when the source is to be used as an entangled photon pair source as a quality figure of merit. It will be evaluated using the single-photon coupling probability $P_1$ of having at least one photon transmitted by the filters.  In the following subsections, a detailed calculation of each of these parameters is given.

\subsubsection{Source brightness}

The coincidental presence of both photons in the target spatial mode over a whole pulse duration is based on the two-photon state $\ket{\psi_2}$ (Eq. \eqref{eq:CoupledState}) and is given by 
\begin{equation}\label{eq:PAB}
	P_2 = \braket{\psi_2}{\psi_2}.
\end{equation}

To make further calculations, we take the limit of an infinite quantization volume, in which sums over modes $\ell$ are replaced by integrals over the wavevectors $\kk$ in the usual way \cite{Garrison2008,Shapiro2009} with discrete variables like $\kk_\ell$, $\w_\ell$ and operators like $\aao[\w_\ell]$ changed into their continuous counterparts $\kk$, $\w(\kk)$ and $\aao\!\!(\w(\kk))$. Moreover, the narrow bandwidth assumption allows for calculating integrals over $\kkappa$ independently of integrals over $\w$, since $\kkappa$ mostly varies with the wavevector angle when $\w$ is restricted to a small interval. Equation \eqref{eq:PAB} gives, under the previous assumptions
\begin{equation}\label{eq:PABgeneral}
	P_2 = |C|^2 \Omega_2 \K_2,
\end{equation}
where
\begin{equation}\label{eq:fluo:PsiZero}
	C = \j e^{\j (\nfs \wso + \nfi \wio) z_0 /c} \sqrt{ \frac{\energiepompe \chieff^2 \wso \wio}{8 \epsilon_0 c^3\np \ns \ni} } 
\end{equation} 
is a constant with $\nfs$ (resp. $\nfi$) the refractive index in the medium where $\OO_0$ describes the target mode. 

The functions
\begin{multline}\label{eq:Omega2}
	\Omega_2 = \Int[\ws] \Int[\wi] \Big|\Tgtemps(\ws+\wi-\wpo) \\
	\times \FF(\ws-\wso) \FF(\wi-\wio) \Big|^2
\end{multline}
and
\begin{multline}\label{eq:fluo:PsiKappa}
	\K_2 = \bigg| \IInt[\kkappa_s] \IInt[\kkappa_i] \Tgespace(\kkappa_s + \kkappa_i, 0) \\
	\times \TOO_0(\kkappa_s,z_0) \TOO_0(\kkappa_i,z_0) e^{-\j z_0(\frac{\np}{\nfs} \frac{|\kkappa_s|^2 }{\kpo}+\frac{\np}{\nfi}  \frac{|\kkappa_i|^2 }{\kpo}) }\\
\times L \sinc \left( \frac{\Delta K (\kkappa_s,\kkappa_i)\: L}{2} \right) \bigg|^2 
\end{multline}
describe respectively the spectral and spatial dependence of $P_2$. They must be both maximized in order to optimize the brightness.

The function $\Omega_2$ involves the coupling of the temporal dependence of the pump pulse with the spectral filters. It has the dimension of a frequency and can be interpreted as the effective source bandwidth. Using an adequate change of variables, it can be written as a convolution of the three spectrally-dependent functions and, as such, it is maximum when the filters are tuned so as to satisfy the energy conservation $\wso + \wio = \wpo$~:
\begin{equation}\label{eq:omega2_convolution}
	\Omega_2 = | \Tgtemps |^2 \conv | \FF |^2 \conv | \FF |^2 (\wso + \wio - \wpo)
\end{equation}
In the following, we will assume that this energy conservation condition is satisfied.

The dimensionless function $\K_2$ takes into account the spatial interferences caused both by phase matching and coupling to the target mode. Its maximization will require a numerical optimization (see \ref{par:SpatialOptimization}).

The factor $|C|^2 $ appears then as a spectral probability density.

\subsubsection{Pair coupling efficiency}

The pair coupling efficiency $\Gamma_2 $ is the ratio of the spectrally and spatially filtered pair probability (source brightness) to the spectrally filtered pair probability (total brightness)
\begin{equation}\label{eq:CouplingEfficiency}
	\Gamma_2 = \frac{P_2}{P_0}.
\end{equation}

The total brightness is calculated from the spectrally filtered, free space state $\ket{\psi_0}$ defined by Eq. \eqref{eq:SpectrallyFilteredState}
\begin{equation}\label{eq:P0}
	P_0 = \braket{\psi_0}{\psi_0}.
\end{equation}

Using the same method as for $P_2$, Eq. \eqref{eq:P0} leads to
\begin{equation}\label{eq:P0general}
	P_0 = |C|^2 \Omega_2 \K_0,
\end{equation}
where $\K_0$ is a dimensionless function describing the spatial dependence and defined by
\begin{multline}
	\K_0 = \IInt[\kkappa_s] \IInt[\kkappa_i] \bigg| \Tgespace(\kkappa_s + \kkappa_i,0) \\
	\times L \sinc \left( \frac{\Delta K (\kkappa_s,\kkappa_i)\: L}{2} \right) \bigg|^2.
\end{multline}

The pair coupling efficiency is then equal to
\begin{equation}\label{eq:PairCouplingEfficiency}
	\Gamma_2 = \frac{\K_2}{\K_0}.
\end{equation}

\subsubsection{Heralding ratio and single-photon coupling}
 
Evaluating the heralding ratio requires calculating the single-photon coupling probability defined by
\begin{equation}\label{eq:P1}
	P_1= \braket{\psi_1}{\psi_1}
\end{equation}
for the state
\begin{equation}
	\ket{\psi_1} = \sum_{\ell_s,\ell_i} \gamma_0(\ell_s,\ell_i) \gamma_T^{(1)}(\ell_s) \gamma_S^{(1)}(\ell_s) \aco[\w_{\ell_s}] \ac_{\ell_i} \ket{0},
\end{equation}
where
\begin{equation}
	\gamma_T^{(1)} = \FF(\w_{\ell_s}-\wso,\kkappa_{\ell_s})
\end{equation}
corresponds to a spectral filtering of the signal and no filtering for the idler \footnote{The natural phase matching bandwidth has no influence on the results that follow if it is much larger than the pump linewidth.}, and
\begin{equation}
	\gamma_S^{(1)} = \frac{1}{\sqrt{\SS}} \TOO^*_{0}(\kkappa_{\ell_s},z_0) e^{\j k'_{z,\ell_s} z_0}
\end{equation}
describes the coupling of the signal photon only into the target single mode.

In the same way as for $P_2$ and $P_0$, 
\begin{equation}\label{eq:SinglePhotonCouplingEfficiency}
	P_1 = |C|^2  \Omega_1 \K_1,
\end{equation}
where
\begin{align}
	\Omega_1 &= \!\!\!\Int[\ws] \!\!\!\Int[\wi] \!\!\!\left|\Tgtemps(\ws+\wi-\wpo) \FF(\ws-\wso)\right|^2 \label{eq:Omega1} \\
	\K_1 &= \!\!\!\IInt[\kkappa_i] \!\!\!\left| \IInt[\kkappa_s]\Tgespace(\kkappa_s + \kkappa_i,0) \TOO_0(\kkappa_s,z_0) \right.\notag\\
		&\qquad \qquad\left.\times e^{-\j z_0\frac{\np}{\nfs} \frac{|\kkappa_s|^2 }{\kpo} }L \sinc \frac{\Delta K \: L}{2}\right|^2.
\end{align}

As $\K_2$, $\K_1$ requires numerical computation, whereas $\Omega_1$ reduces to
\begin{equation}
	\Omega_1 = \intw |\FF(\w)|^2
\end{equation}
which is proportional to the filter bandwidth itself.

The heralding ratio i.e. the conditionnal probability to couple the second photon to the target mode when its twin has been coupled, is defined by
\begin{equation}\label{eq:HeraldingRatio}
	\Gamma_{2|1} = \frac{\K_2}{\K_1}.
\end{equation}

It is useful to define also the single photon coupling efficiency
\begin{equation}\label{eq:SinglePhotonCoupling}
	\Gamma_{1} = \frac{\K_1}{\K_0}.
\end{equation}

These two coefficients are related to the pair coupling efficiency in the following way
\begin{equation}\label{eq:ProbaCondi}
	\Gamma_{2} = \Gamma_{2|1} \: \Gamma_{1}.
\end{equation}

Note that $\Gamma_{2} \neq \Gamma_{1}^2$, because of the spatial correlation between the two photons, in the same way as, because of the energy correlation of the two photons, the calculation of $\Omega_2$  involves a convolution rather than a product of the filtering functions of the signal and idler photons (see Eq. \eqref{eq:omega2_convolution}).

In order to get a physical insight into these results and show how this description can be used as a predictive tool for the design and optimization of entangled photon pair sources, the following section is devoted to a numerical calculation corresponding to the particular case of the experiment described in Section \ref{sec:Experiment}.

\section{Numerical optimization of a narrow-band fibered source pumped by a Gaussian beam}\label{sec:NarrowBand}

\subsection{Gaussian pump beam, Gaussian target mode}

We assume the pump beam to be Gaussian with a waist radius $w_0$ and Rayleigh length $z_R$
\begin{align}
		\gespace(\rrho,0) &= \sqrt{\frac{2}{\pi w_0^2}} \: e^{-\abs{\rrho}^2/w_0^2} \label{eq:EnvelopeSpaceGauss}\\
	\Tgespace(\kkappa,0) &= \sqrt{2\pi w_0^2} \: e^{-\abs{\kkappa}^2 w_0^2 / 4}.
\end{align}

Using the fact that $\kpo/2 = z_R/w_0^2$, the phase matching function becomes
\begin{multline}\label{eq:DeltaKgauss}
	\Delta K\approx \Delta {k_0} + \frac{w_0^2}{4 z_R} \bigg[ |\kkappa_s + \kkappa_i|^2 \\
		- 2 \Big( 1 - \frac{\delta \w}{\wpo} \Big) \frac{\np}{\ns} |\kkappa_s|^2
		- 2 \Big( 1 + \frac{\delta \w}{\wpo} \Big) \frac{\np}{\ni} |\kkappa_i|^2 \bigg].
\end{multline}

The target mode acting as a spatial filter is defined by the profile of a Gaussian mode at its waist of radius $a_0$ located at $z_0$. This can be the transverse mode of a fiber or its image by a lens collection system
\begin{align}
\OO_0(\rrho,z_0) &= \sqrt{\frac{2}{\pi a_0^2}} \: e^{-|\rrho|^2/a_0^2}\\
	\TOO_0(\kkappa,z_0) &= \sqrt{2 \pi a_0^2} \: e^{-|\kkappa|^2 a_0^2 / 4}.
\end{align}

\subsection{Nondimensionalization}

The coincidence probability $P_2$ given by Eq. \eqref{eq:PABgeneral} can now be detailed. It is useful to separate the fixed parameters from the configuration of the experiment that can be optimized
\begin{equation}\label{eq:PABparticular}
	P_2 = \frac{\energiepompe \chieff^2 L \DwF \wso \wio \wpo}{8 \epsilon_0 c^4 \ns \ni} \cdot \!\frac{\Omega_2 }{\DwF}(\vdelta) \cdot \!\frac{\K_2}{\kpo L}(\veta, \vrho, \vzeta, \vphiz),
\end{equation}
where $\DwF $ is the filter bandwidth, and the terms $\Omega_2 / \DwF$ and $\K_2/(\kpo L)$ depend only on the experimental configuration described by the following dimensionless parameters :
\begin{subequations}
	\begin{align*}
	  \vdelta  &= \frac{4 \ln 2}{\Dtp \DwF } && \text{\footnotesize{}(relative pump bandwidth)}\\
	  \veta    &= \frac{L}{2 z_R} && \text{\footnotesize{}(pump focusing parameter)}\\
	  \vrho    &= \frac{a_0}{w_0} && \text{\footnotesize{}(normalized target mode waist size)}\\
	  \vzeta   &= \frac{z_0}{L} && \text{\footnotesize{}(longitudinal collection offset)}\\
	  \vphiz   &= \Delta k_0 \, L \phantom{\frac{L}{L}} && \text{\footnotesize{}(longitudinal phase mismatch)}\\
	  \vvphis  &= \frac{w_0}{2} \kkappa_s && \text{\footnotesize{}(normalized signal transverse wavevector)}\\
	  \vvphii  &= \frac{w_0}{2} \kkappa_i && \text{\footnotesize{}(normalized idler transverse wavevector)}
	\end{align*}
\end{subequations}

The parameter $\Dtp$ is the pump pulse duration. Eq. \eqref{eq:PABparticular}, that will be used to calculate the pair coupling efficiency and the heralding ratio, has a particular importance in itself, as it quantifies the absolute coupled brightness and its dependence on various experimental parameters. It will be discussed and compared to other reported work in Sec. \ref{sec:Discussion}. The following section is dedicated to the theoretical determination of the experimental configuration that maximizes this brightness.

\subsection{Optimization of the source brightness}

\subsubsection{Spectral optimization}

The optimisation of Eq. \eqref{eq:Omega2} must generally be made numerically, but the influence of the relative pump bandwidth $\vdelta$ can be developped in a fully analytical way if the pump temporal envelope and the spectral filters are both Gaussian
\begin{align}
	\gtemps(t) &= \left[ \frac{4 \ln 2}{\pi\Dtp^2} \right]^{\frac{1}{4}} e^{- 2 \ln 2 \, t^2 / \Dtp^2} \label{eq:EnvelopeTempGauss} \\	
	\Tgtemps(\w) &= \left[ \frac{\pi\Dtp^2}{\ln 2} \right]^{\frac{1}{4}} e^{ -\w^2 \Dtp^2 / (8 \ln 2) }\\
	\FF(\w) &= e^{- 2 \ln 2 \, \w^2 / \DwF^2}\label{eq:fluo:FormeFiltreSpectral},
\end{align}
where  $\Dtp$ and $\DwF$ are full widths at half maximum intensity. The coincidence probability then depends on $\delta$ in the following way
\begin{equation}
	\frac{\Omega_2 }{\DwF}(\vdelta) = \sqrt{ \frac{ \pi/(8\ln 2) }{1 + \frac{\vdelta^2}{2}}} 
	\label{eq:fluo:Ftemps}
\end{equation}

\begin{figure}
	\centering
	\includegraphics[width=\columnwidth]{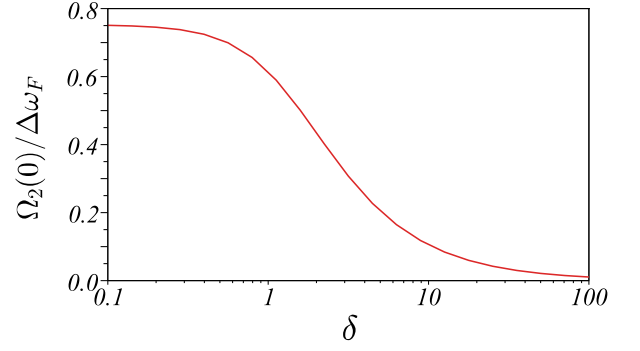}
	\caption{Effect of the relative pump linewidth $\vdelta$ on the spectral transmission factor $\Omega_2 /\DwF$ }
	\label{fig:TransmissionFactorSpectral}
\end{figure}

This dependence of the spectral factor of the source brightness is plotted in Figure \ref{fig:TransmissionFactorSpectral} as a function of the relative pump bandwidth. It shows that the maximum value is asymptotically reached when the pump beam is monochromatic. In this case, the joint probability for a photon and its twin to be transmitted is optimal. When the pump linewidth increases, some idler photons at $\wi = \wp - \ws$ and their corresponding signal photon at $\ws$ are not frequency-symmetric with respect to the filter center frequency ($\wp \neq \wpo$), which is necessarily detrimental. This effect becomes significant for $\Dwp \geq \DwF$ ($\vdelta \geq 1$). Although reducing the pulse duration $\Dtp$ in order to increase the repetition rate does increase the effective photon pair rate, it is desirable to compromise in order to keep $\Omega_2$ close to its maximum value.

\subsubsection{Spatial optimization for degenerate down-conversion}\label{par:SpatialOptimization}

In the following, we restrict ourselves to the case of a frequency-degenerate down-conversion ($\ws=\wi$), which corresponds to the experimental setup described in section \ref{sec:Experiment}. The dimensionless term to be optimized has the following expression
\begin{equation}\label{eq:K2}
	\!\!\!\!\frac{\K_2}{\kpo L}(\veta, \vrho, \vzeta, \vphiz) = \frac{8}{\pi^5} \veta \vrho^4 \Big| \iint \! d^2\vvphis \iint \! d^2\vvphii \; \Q_2 \Big|^2
\end{equation}
with
\begin{align}
\Q_2 &= \exp \Big\{ - | \vvphis + \vvphii |^2 \Big\} \label{eq:IntegrandPAB} \\
	&\times \exp \Big\{- \vrho^2 \big( | \vvphis |^2 + | \vvphii |^2 \big) \Big\} \notag \\
	&\times \exp \: \j \Big\{- 4 \veta \vzeta \left( \frac{\np}{\nfs} | \vvphis |^2 + \frac{\np}{\nfi} | \vvphii |^2 \right) \Big\} \notag \\
	&\times \sinc \Big\{ \frac{\vphiz}{2} + \veta \Big[ | \vvphis + \vvphii |^2  - 2 \frac{\np}{\ns} |\vvphis|^2 - 2 \frac{\np}{\ni} |\vvphii|^2 \Big] \Big\} \notag
\end{align}

Using polar coordinates, we can use the following mapping~:
\begin{multline}\label{eq:3Dintegral}
	\iint \! d^2\vvphis \iint \! d^2\vvphii \; \Q_2 \\
	\longrightarrow 2 \pi \int_0^\infty \!\!\!\! \rhos d\rhos \int_0^\infty \!\!\!\! \rhoi d\rhoi \int_0^{2\pi} \!\!\!\! d(\thetas-\thetai) \: \Q'_2
\end{multline},
with
\begin{align}
	\Q'_2 &= \exp \Big\{ - \big( 1 + \vrho^2 \big) \big( \rhos^2 + \rhoi^2 \big) \Big\} \notag \\
	&\times \exp \Big\{ - 2 \rhos \rhoi \cos(\thetas - \thetai) \Big\} \notag \\
	&\times \exp \: \j \Big\{- 4 \veta \vzeta \left( \frac{\np}{\nfs} \rhos^2 + \frac{\np}{\nfi} \rhoi^2 \right) \Big\} \notag \\
	&\times \sinc \Big\{ \frac{\vphiz}{2} + \veta \Big[ \big( 1-2\frac{\np}{\ns} \big) \rhos^2 + \big( 1-2\frac{\np}{\ns} \big) \rhoi^2 \notag \\
	&\qquad \qquad + 2 \rhos \rhoi \cos(\thetas - \thetai) \Big] \Big\}.
\end{align}

In this way, the quadruple integral turns into a triple integral which is numerically evaluated using an adaptative 3D quadrature algorithm \cite{Berntsen1991}.

\begin{figure}
	\centering
	\includegraphics[width=1.0\columnwidth]{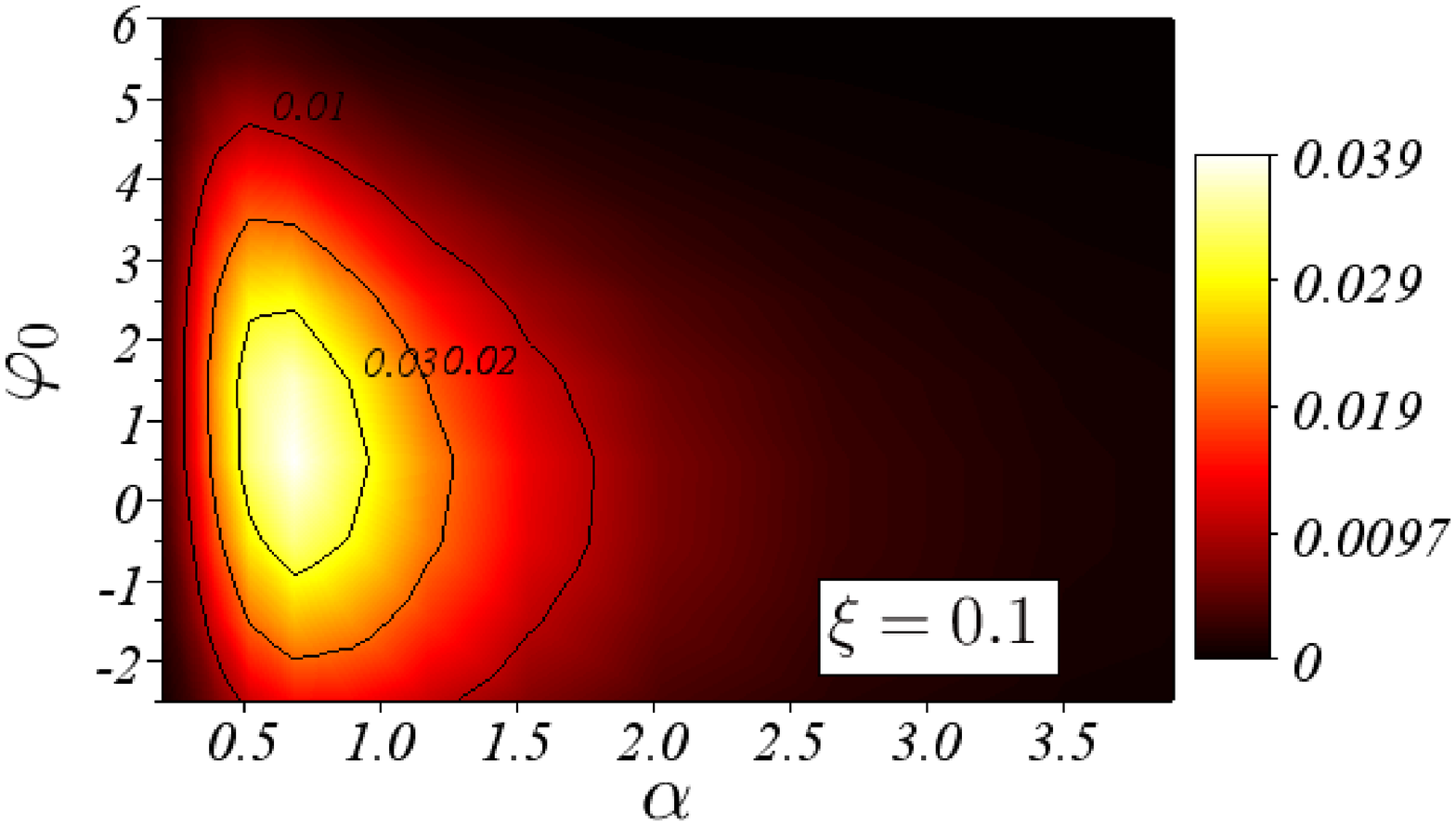}\\
	\includegraphics[width=1.0\columnwidth]{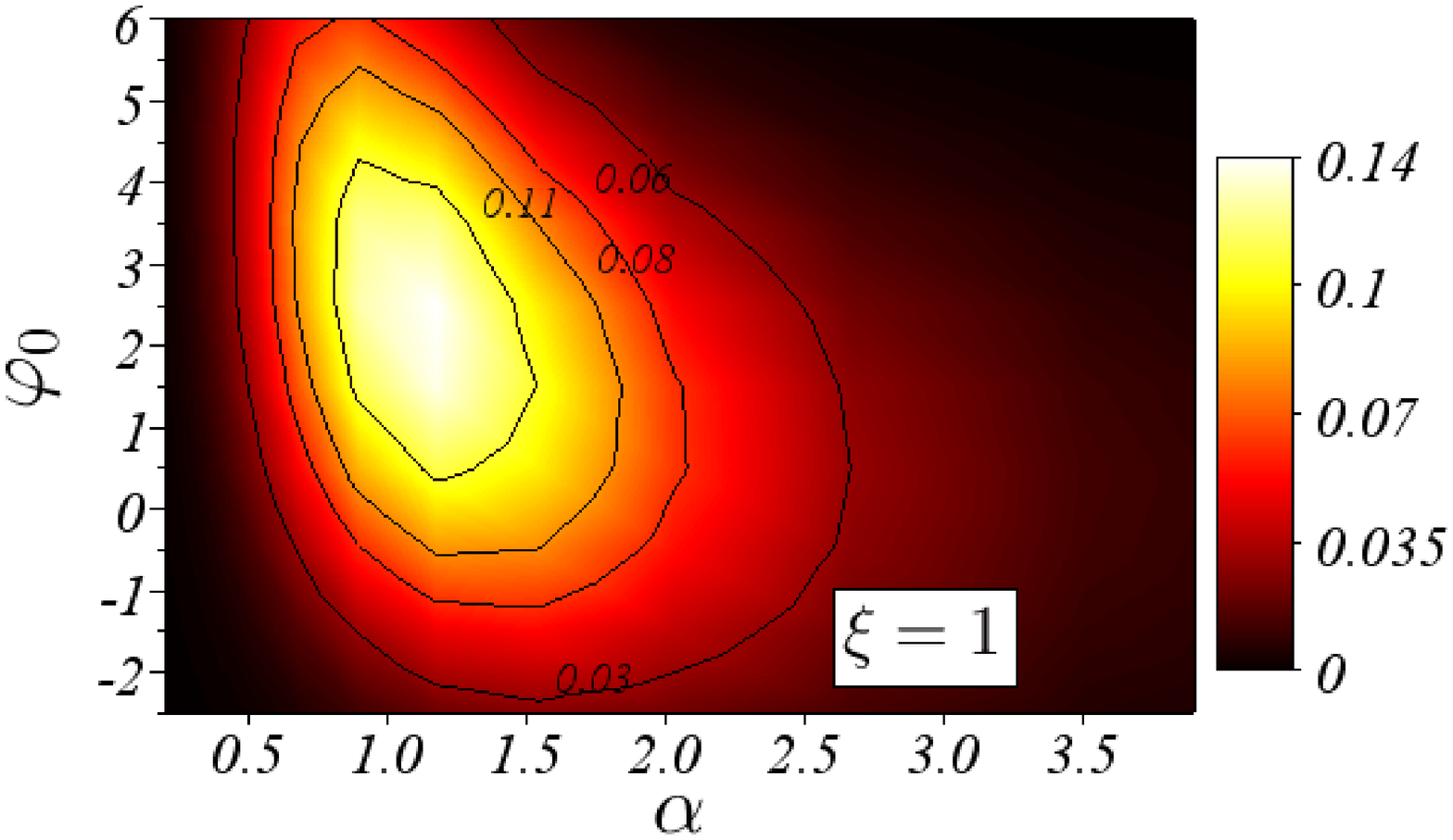}\\
	\includegraphics[width=1.0\columnwidth]{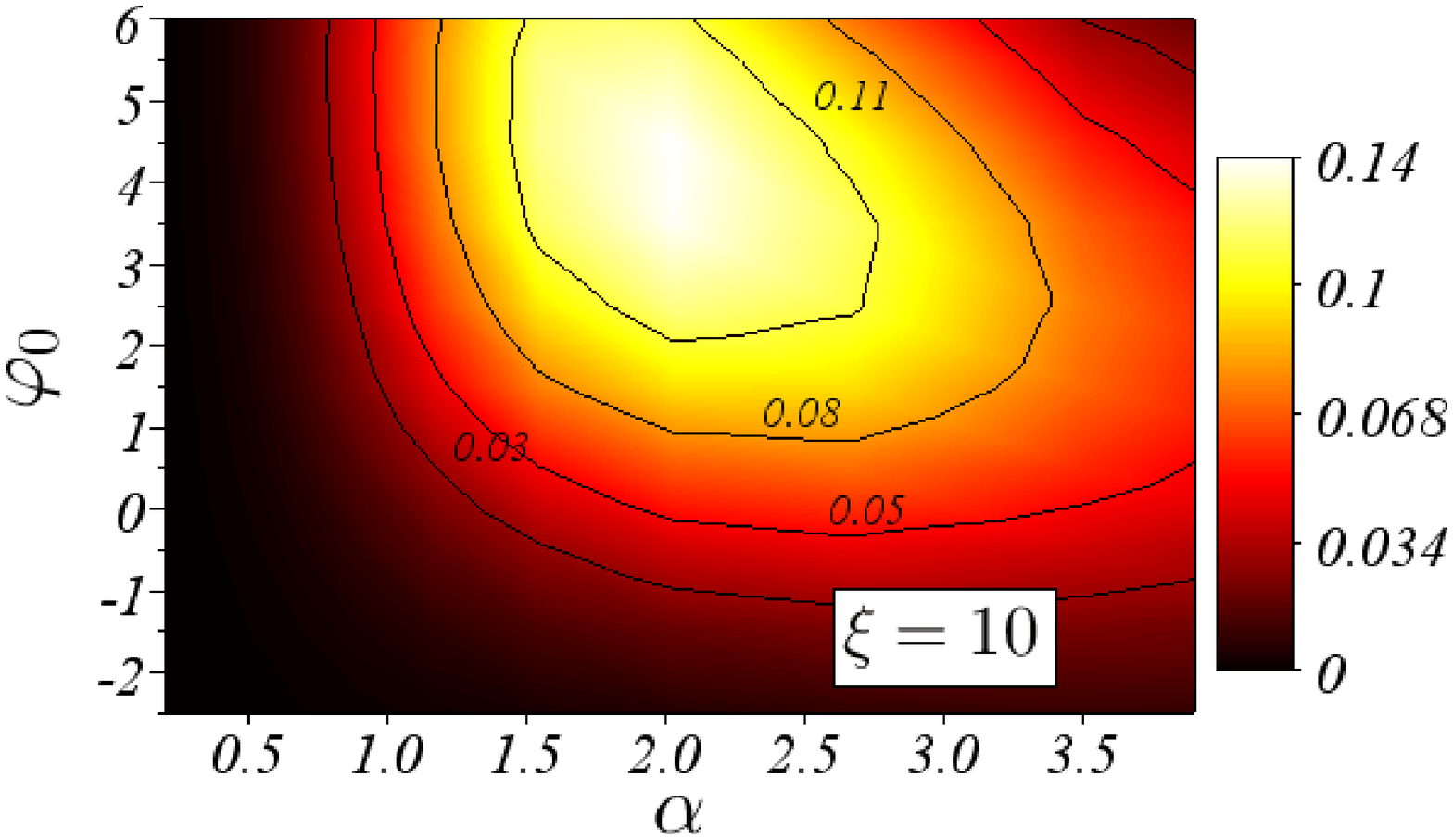}
	\caption{Optimization of the spatial filtering factor $\K_2/(\kpo L)$ with respect to the normalized target mode waist $\vrho$ and the longitudinal phase mismatch $\vphiz$ for three values of the focusing parameter ~: $\veta = 0.1$ (a) , $\veta= 1$ (b) , $\veta =10$ (c) .} 
	\label{fig:TransmissionFactorSpatialOptimization}
\end{figure}

\begin{figure}
	\centering
	\includegraphics[width=\columnwidth]{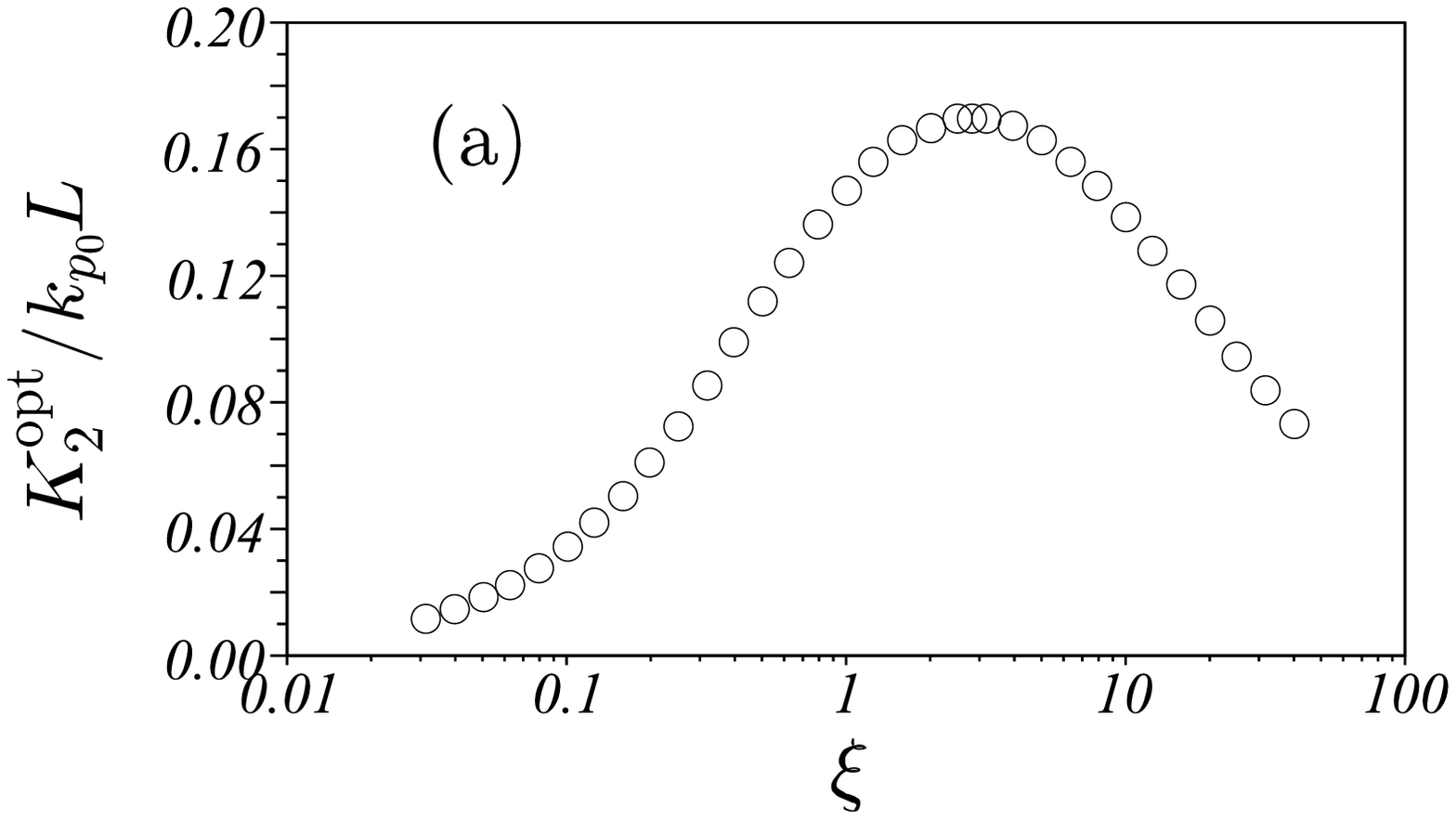}\\
	\includegraphics[width=\columnwidth]{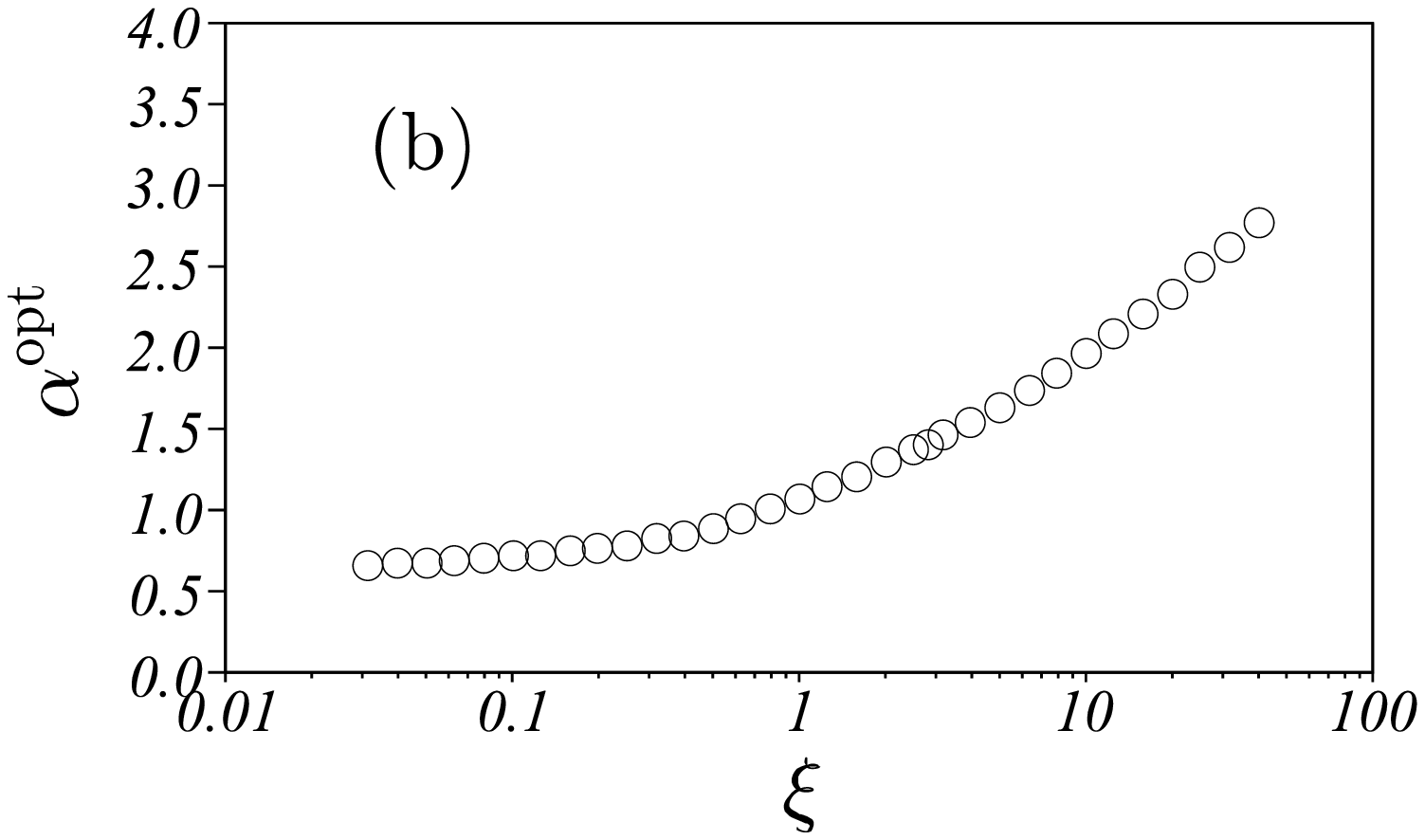}\\
	\includegraphics[width=\columnwidth]{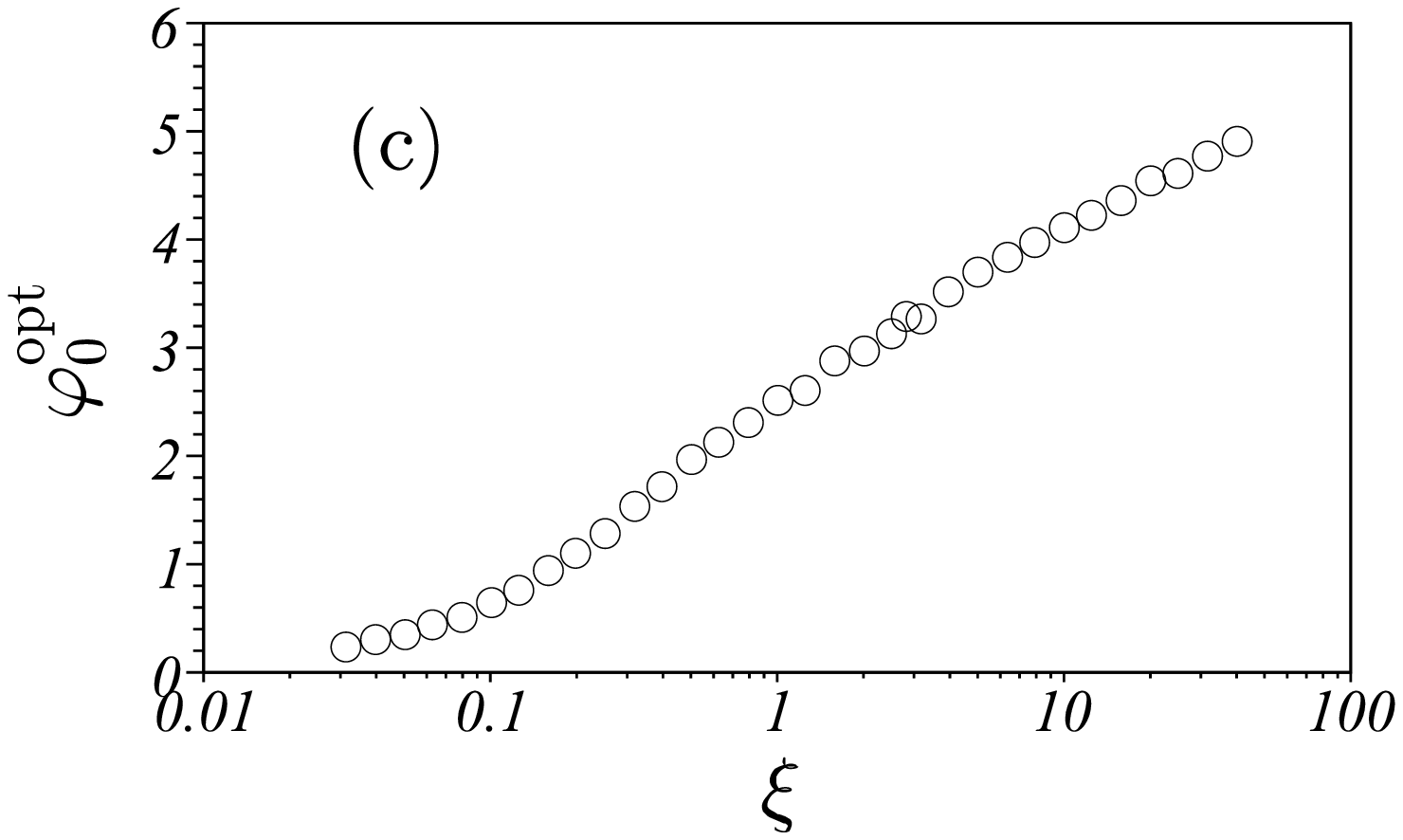}
	\caption{Maximal value of the spatial filtering factor $\K_2/(\kpo L)$ (a) and associated optimal values of the normalized target mode waist $\vrho$ (b) and the longitudinal phase mismatch $\vphiz$ (c), for various values of the focusing parameter $\veta$. }
	\label{fig:TransmissionFactorSpatialOptimal}
\end{figure}

Figure \ref{fig:TransmissionFactorSpatialOptimization} represents, for three different focusing parameters $\veta$, the value of the spatial filtering term $\K_2/(\kpo L)$ as a function of parameters $\vrho$ and $\vphiz$. It shows that there exists a unique couple $(\vrho^\text{opt},\vphiz^\text{opt})$ that allows reaching the maximum $\K_2^\text{opt}/(\kpo L)$ for a given value of $\veta$. Moreover, that maximum varies with $\veta$ and the accuracy of $(\vrho^\text{opt},\vphiz^\text{opt})$ is found to be more critical for low $\veta$. Indeed, when the focusing of the pump beam increases, phase matching can only be satisfied in an average way, because of the large range of emitted angles, and $\vrho^\text{opt}$ results from a compromise between collecting weakly divergent photons with high efficiency and reducing this efficiency to collect more strongly divergent photons. Optimizing the pump beam for photon pair collection indeed consists in finding the focusing parameter for which these two compromises offer the best performance.

Note that in Figure \ref{fig:TransmissionFactorSpatialOptimization}, the normalized longitudinal offset of single-mode collection $\vzeta$ has been set to zero. We have checked that this choice gives optimal results. The optimality of $\vzeta=0$ is due to the symmetry of the problem in our particular choice of a Gaussian pump beam and Gaussian target spatial mode ; this might not be the case in other circumstances.

In order to find the focusing parameter $\veta$ that gives rise to the highest collected brightness $\K_2^\text{opt}/(\kpo L)$, we have performed optimizations similar to that of Fig. \ref{fig:TransmissionFactorSpatialOptimization} for values of $\veta$ ranging from $0.03$ to $40$, that is for waist radiuses from $\sim 200$ to $\sim 5$ $\mathrm{\mu}$m for a red pump in common crystals, for instance. Knowing that $\vzeta=0$ is optimal in the whole range, we have plotted on Fig. \ref{fig:TransmissionFactorSpatialOptimal} the values $\K_2^\text{opt}/(\kpo L)$
, $\vrho^\text{opt}$ 
 and $\vphiz^\text{opt}$ 
as a function of $\veta$. 

According to Fig. \ref{fig:TransmissionFactorSpatialOptimal}b, the optimum normalized target mode waist $\vrho^\text{opt}$ does not vary very much over this large range of focusing parameters. Starting with an approximate matching of the target mode to the pump waists $\vrho^\text{opt} \approx 1$ at low focusing, our calculations exhibit a slow increase, showing that when focusing gets stronger, it is preferable to collect the weakly divergent photon pairs, since modes of larger waist sizes have smaller numerical apertures. On Fig. \ref{fig:TransmissionFactorSpatialOptimal}c, the optimum longitudinal phase mismatch $\vphiz^\text{opt}$ increases with the focusing parameter, as to compensate for the transverse mismatch caused by the strong focusing.

Fig. \ref{fig:TransmissionFactorSpatialOptimal}a shows that the value of the focusing parameter giving the highest value of $\K_2^\text{opt}/(\kpo L)$ is $\veta=2.84$, for which the longituninal phase mismatch is $\vphiz=3.2$. These values correspond to the Boyd and Kleinman conditions \cite{Boyd1968} for a highest second harmonic generation efficiency with Gaussian beams. Indeed, when evaluating the brightness produced in a single Gaussian mode, the collinear degenerate down-conversion is a process symmetric to second harmonic generation.

However, down-converted photons are not entirely produced in a single mode. That is why it is important to investigate the optimization of the pair coupling efficiency $\Gamma_2$ defined in Eq. \eqref{eq:CouplingEfficiency}.

\subsection{Optimization of the pair coupling efficiency}

The total source brightness must also be calculated, using the dimensionless variables defined previously~:
\begin{equation}\label{eq:P0particular}
	P_0 = \frac{\energiepompe \chieff^2 L \DwF \wso \wio \wpo}{8 \epsilon_0 c^4 \ns \ni} \cdot \frac{\Omega_2 }{\DwF}(\vdelta)\cdot \frac{\K_0}{\kpo L}(\veta, \vphiz),
\end{equation}
where
\begin{equation}\label{eq:K0}
	\frac{\K_0}{\kpo L} = \frac{2}{\pi^3} \veta \iint\!d^2\vvphis \iint\!d^2\vvphii \; \left| \Q_0 \right|^2,
\end{equation}
with
\begin{align}
	\Q_0 &= \exp \Big\{ - | \vvphis + \vvphii |^2 \Big\} \label{eq:IntegrandP0} \\
	&\times \sinc \Big\{ \frac{\vphiz}{2} - \veta \Big[ | \vvphis + \vvphii |^2  - 2 \frac{\np}{\ns} |\vvphis|^2 - 2 \frac{\np}{\ni} |\vvphii|^2 \Big] \Big\} \notag
\end{align}
which can also be reduced to a 3D-integral.

\begin{figure}
	\centering
	\includegraphics[width=\columnwidth]{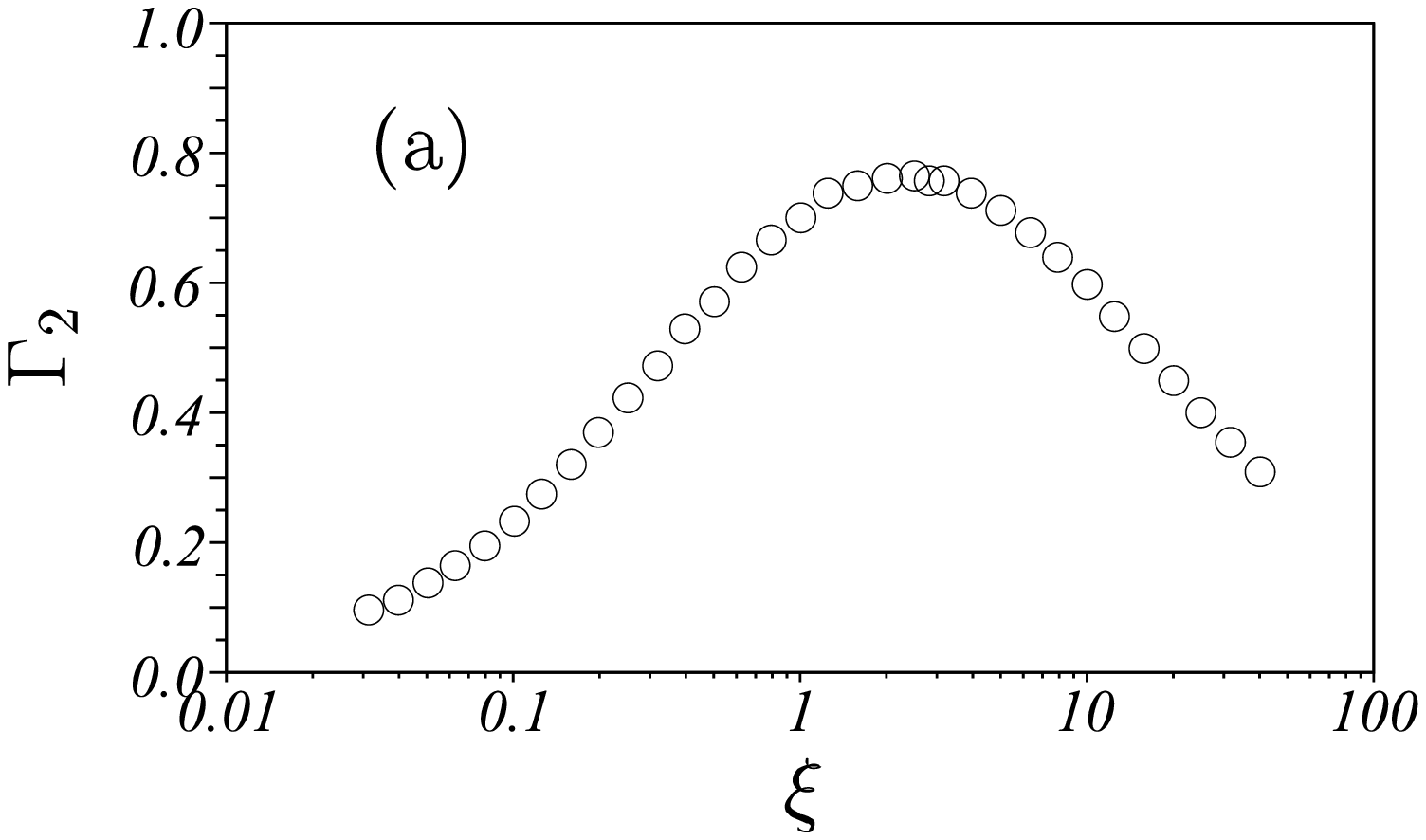}\\
	\includegraphics[width=\columnwidth]{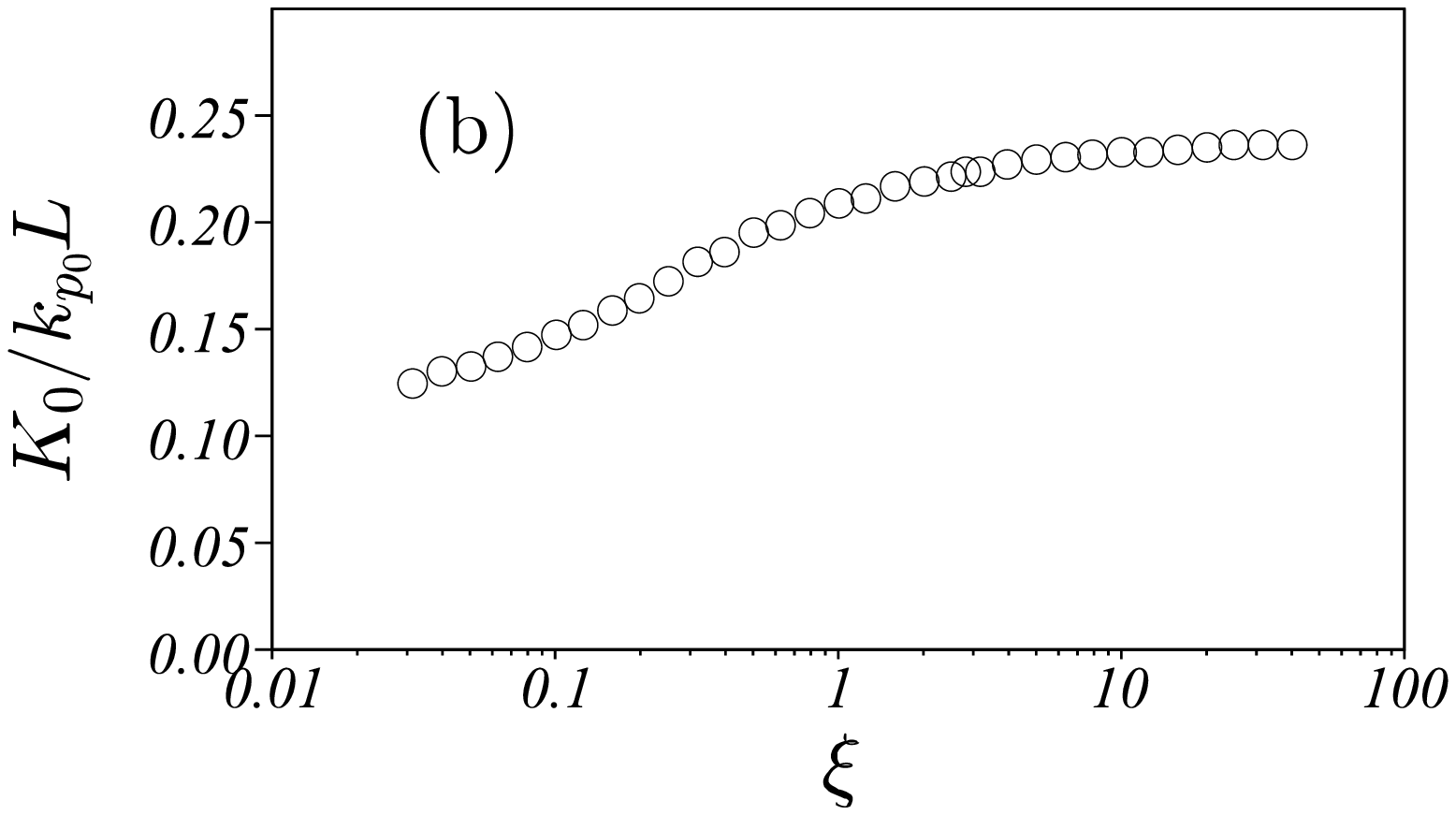}
	\caption{Pair coupling efficiency $\Gamma_2$ (a) compared to the spatial factor $\K_0/(\kpo L)$ of the pair production probability (b), for various values of the focusing parameter $\veta$. $\K_0/(\kpo L)$ is evaluated for the longitunal phase mismatch $\vphiz$ that maximizes $\Gamma_2$.}
\label{fig:CouplingVsFSHC}
\end{figure}

The pair coupling efficiency $\Gamma_2$ (Eq. \eqref{eq:PairCouplingEfficiency}) is calculated using Eqs. \eqref{eq:K2} and \eqref{eq:K0}. It is plotted in Figure \ref{fig:CouplingVsFSHC}a, along with the dimensionless $\K_0/(\kpo L)$ proportional to the total brightness $P_0$ (Fig. \ref{fig:CouplingVsFSHC}b), as a function of the focusing parameter $\veta$. The total brightness does not vary much for $\veta \gtrsim 1$, therefore the coupling efficiency follows the same tendency as the single-mode brightness (Fig. \ref{fig:TransmissionFactorSpatialOptimal}a). However, the precise optimal value of $\veta$ is no longer $\veta = 2.84$ but is closer to $2$. For a red pump in common crystals, this means a waist radius difference of $20$ \%. Under different assumptions from the ones considered in this section, however, the difference might be greater. When power efficiency is critical, as for sources intended to be compact, it can be useful to have in mind that to collect most of the generated photons, optimizing on the mere source brightness is not optimal with respect to pump power consumption.

\subsection{Optimization of the heralding ratio}

The heralding ratio $\Gamma_{2|1}$ (Eq. \eqref{eq:HeraldingRatio}) requires the computation of the single-photon coupling probability $P_1$, (cf. Eq. \eqref{eq:SinglePhotonCouplingEfficiency})~:
\begin{equation}\label{eq:P1particular}
	P_1 = \frac{\energiepompe \chieff^2 L \DwF \wso \wio \wpo}{8 \epsilon_0 c^4\ns \ni} \cdot \!\frac{\Omega_1}{\DwF}(\vdelta) \cdot \!\frac{\K_1}{\kpo L}(\veta, \vrho, \vzeta, \vphiz),
\end{equation}
where $\frac{\Omega_1}{\DwF}$ only depends on the shape of the filter itself and where
\begin{equation}\label{eq:K1}
	\frac{\K_1}{\kpo L} = \frac{4}{\pi^4} \veta \vrho^2 \iint\!d^2\vvphis \left| \iint\!d^2\vvphii \Q_1 \right|^2,
\end{equation}
with
\begin{align}
	\Q_1 &= \exp \Big\{ - | \vvphis + \vvphii |^2 \Big\} \label{eq:IntegrandP1} \\
	&\times \exp \Big\{- \vrho^2 | \vvphis |^2 \Big\} \notag \\
	&\times \exp \: \j \Big\{- 4 \veta \vzeta \frac{\np}{\nfs}| \vvphis |^2 \Big\} \notag \\
	&\times \sinc \Big\{ \frac{\vphiz}{2} - \veta \Big[ | \vvphis + \vvphii |^2  - 2 \frac{\np}{\ns} |\vvphis|^2 - 2 \frac{\np}{\ni} |\vvphii|^2 \Big] \Big\} \notag
\end{align}
which can be reduced to a 3D-integral as for $\Q_2$ and $\Q_0$.

\begin{figure}
	\centering
	\includegraphics[width=1.0\columnwidth]{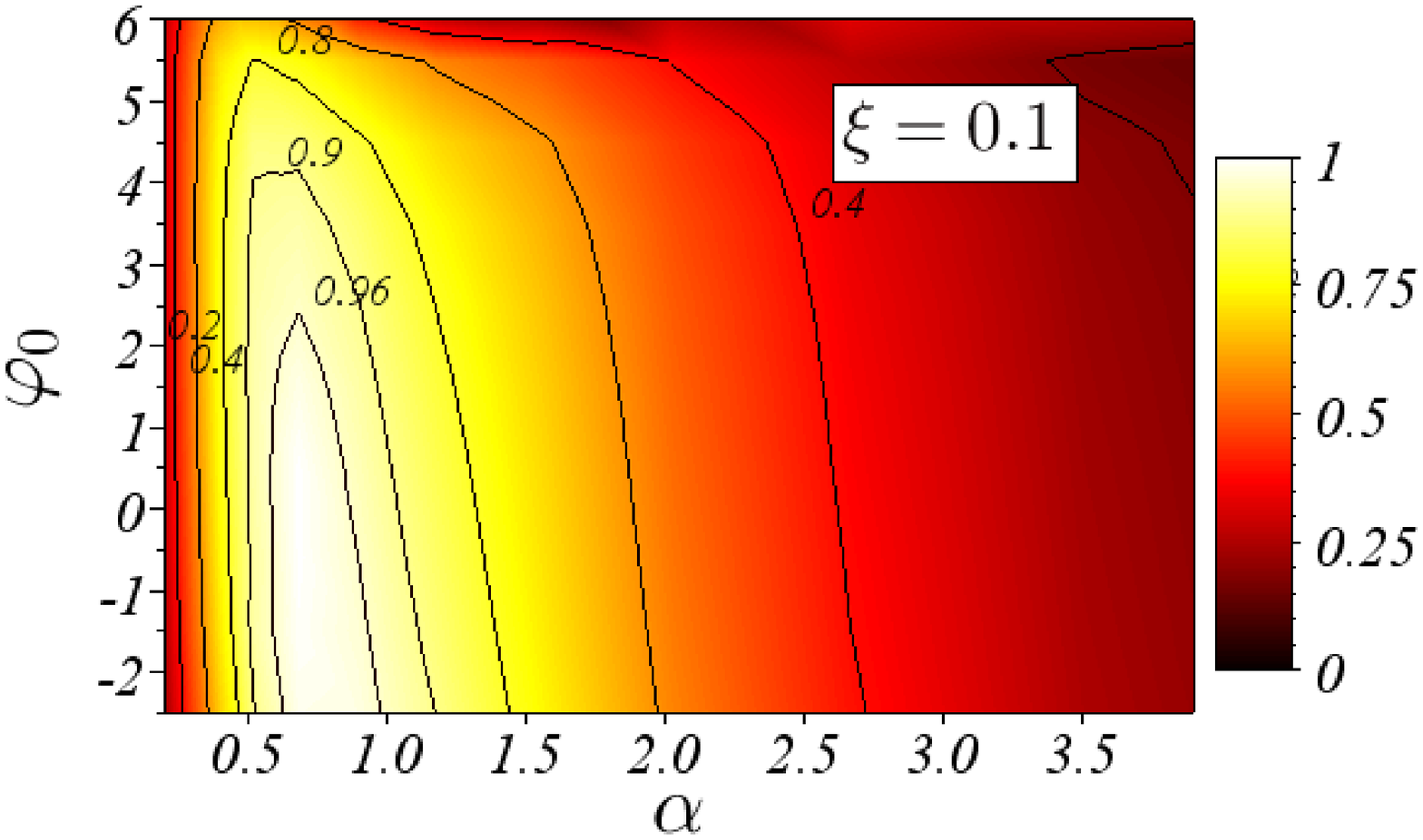}\\
	\includegraphics[width=1.0\columnwidth]{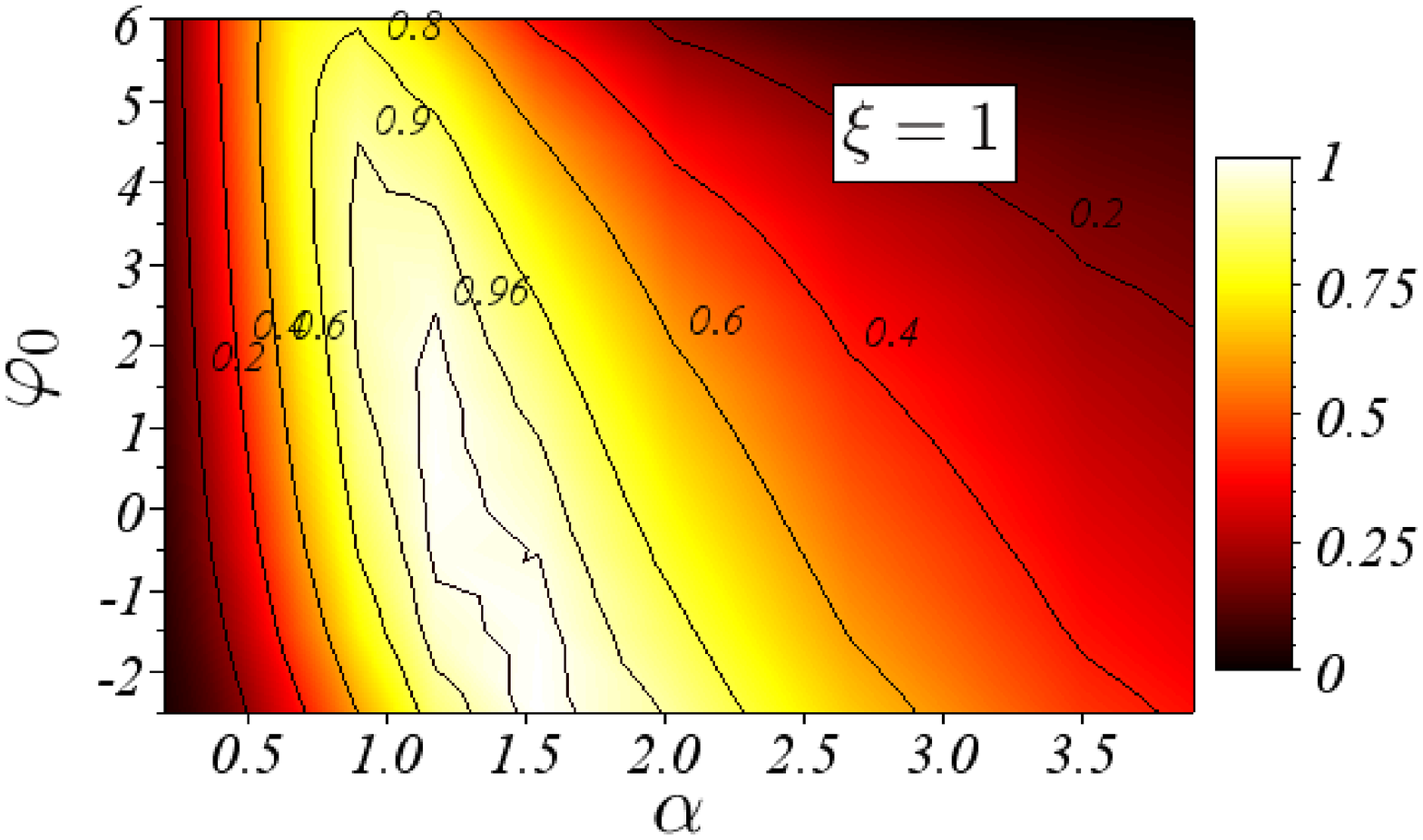}\\
	\includegraphics[width=1.0\columnwidth]{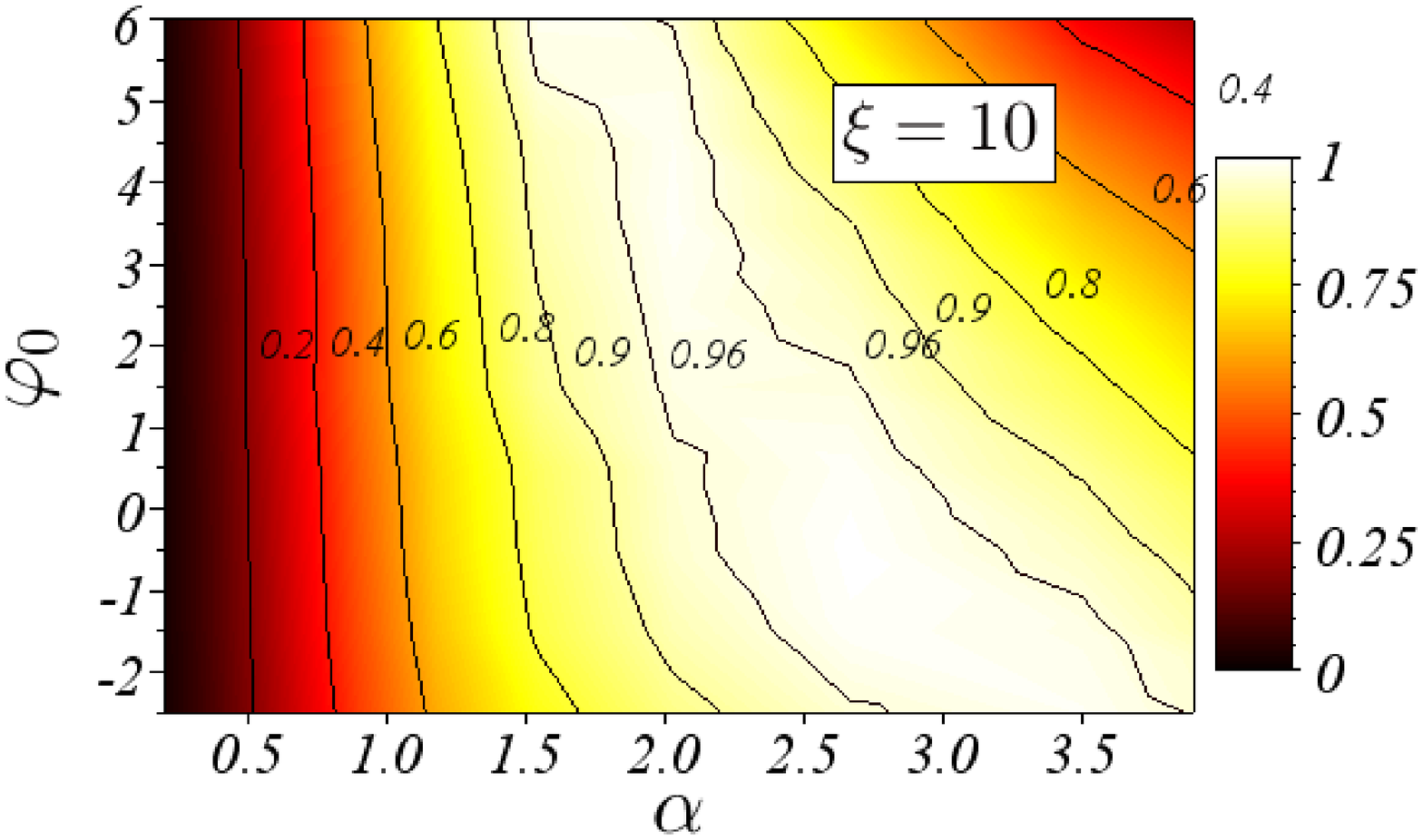}
	\caption{Optimization of the heralding ratio $\Gamma_{2|1}$ with respect to the normalized target mode waist $\vrho$ and the longitudinal phase mismatch $\vphiz$ for three values of the focusing parameter~: $\veta = 0.1$ (a) , $\veta= 1$ (b),  $\veta = 10$ (c).}
	\label{fig:HeraldingRatio}
\end{figure}

The heralding ratio $\Gamma_{2|1} = \K_2 / \K_1$ is plotted in Fig. \ref{fig:HeraldingRatio} as a function of parameters $\vrho$ and $\vphiz$ for three values of the focusing parameter $\veta = 0.1, 1, 10$. Contrary to the brightness depicted in Fig \ref{fig:TransmissionFactorSpatialOptimization}, the heralding ratio can reach a value close to 1 on the whole range of $\veta$. A large range of $\vphiz$ is compatible with this maximum, but the overlap with the range leading to a high brightness is small. On the contrary, the tolerance on $\vrho$ is relatively low. When the results of $P_2$ and $\Gamma_{2|1}$ are both taken into account, the theory gives indeed useful information about the target mode waist for which the collection should be optimized with respect to $\vphiz$ (through the crystal temperature), to find the best compromise between the brightness and the heralding ratio.

The details of an experiment that enabled a validation of our model is given in the following section.

\section{Experimental validation}\label{sec:Experiment}

\subsection{Experimental setup}

\begin{figure}
	\centering
	\includegraphics[width=\columnwidth]{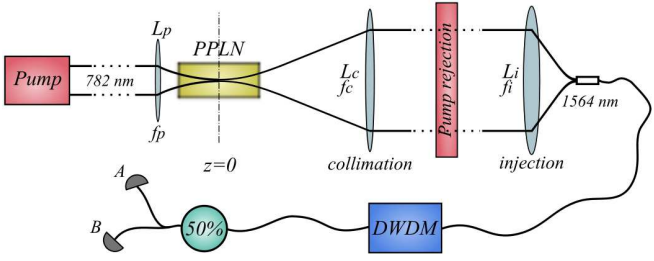}
	\caption{Experimental setup : A pulsed pump laser at wavelength 782 nm is focused in a periodically-poled lithium niobate crystal (PPLN) with a lens $L_p$ of focal length $f_p$. Down-converted photons at 1564 nm are coupled into an single mode fiber through an optical system composed of lenses $L_c$ and $L_i$ of respective focal lengths $f_c$ and $f_i$. Spectral filtering is performed via a DWDM add-drop filter of bandwidth 75 GHz, and photons are split with 50 \% efficiency towards detectors A and B using a balanced fibered coupler.}
	\label{fig:ExperimentalSetup}
\end{figure}

The experimental setup used to validate the theory is depicted in Fig. \ref{fig:ExperimentalSetup}.
SPDC is generated by focusing a pulsed pump beam at 782 nm in a periodically poled lithium niobate (PPLN) crystal of length $L=2$ cm with a poling period $\Lambda = 19.34 \mathrm{µm}$. The mean pump power is $5$ mW and the $25$ ns Gaussian pulses (FWHM) are Fourier transform limited with a $2$ MHz repetition rate. The spatial profile of the pump beam is also Gaussian. On-axis fluorescence around $1564$ nm is collected into a telecom optical fiber through the lenses $L_c$ (achromatic doublet) and $L_i$ (asphere). The same low bandwidth filter ($\DwF=2 \pi \times 75$ GHz) is used for both signal and idler photons so that the source is operated at the degeneracy frequency ($\wso=\wio=\wpo/2$). A balanced coupler is used to split photon pairs with 50 \% efficiency and the photons are detected on paths A and B.

\subsection{Experimental method}

In order to explain how the heralding ratio $\Gamma_{2|1}$ can be determined experimentally, let us show its relation to measured parameters~:
\begin{equation}
	\Gamma_{2|1} = \frac{K_2}{K_1} =\frac{\Omega_1}{\Omega_2} \frac{P_2}{P_1}.
\end{equation}
The experimental parameter $\PAB$ corresponding to the calculated $P_2$  is the measured coincidence probability per pulse from which accidental and noise coincidences are substracted, and $P_1$ is related to the measured counts on channel I from which dark counts have been substracted~:
\begin{equation}
	\PAB =\mathcal{T}_A\mathcal{T}_B P_2 \qquad P_I =\mathcal{T}_I P_1
\end{equation}
where $\mathcal{T}_I$ is the overall tranmission of channel $I$. We obtain~:
\begin{equation}
	\Gamma_{2|1} = \frac{\Omega_1}{\Omega_2} \frac{\PAB}{\mathcal{T}_A \mathcal{T}_B}\frac{\mathcal{T}_I}{P_1}
\end{equation}
The heralding ratio $\Gamma_{2|1}$ can hence be determined from the measurements of counts and coincidences, provided the insertion losses have been previously determined \cite{Smirr2010b}. It is then possible to validate its dependence with respect to the pump focusing parameter $\xi$ and the normalized target mode waist $\alpha$.

The variation of $\xi$ was obtained by changing the lens $L_p$ focusing the pump beam into the PPLN crystal. For each value of the focal length $f_p$, the waist of the pump beam was measured, allowing the determination of $z_R$ and $\veta=L/(2z_R)$.

Counts and coincidences were then measured using various focal lengths $f_i$ of the lens focusing the SPDC beam into the fiber. The value $a_0$ of the image of the fiber waist in the crystal was determined using the magnification factor $f_c/f_i$ of the collection system (composed of the lenses $L_c$ and $L_i$) in order to determine $\vrho=a_0/w_0$. Let us note that each data point requires changing the focusing lens, realigning the setup, and successively optimizing the collection with at least five different injection lenses, with the phase mismatch (crystal temperature) as an additionnal degree of freedom.

\subsection{Experimental results}

\begin{figure}
	\centering
	\includegraphics[width=\columnwidth]{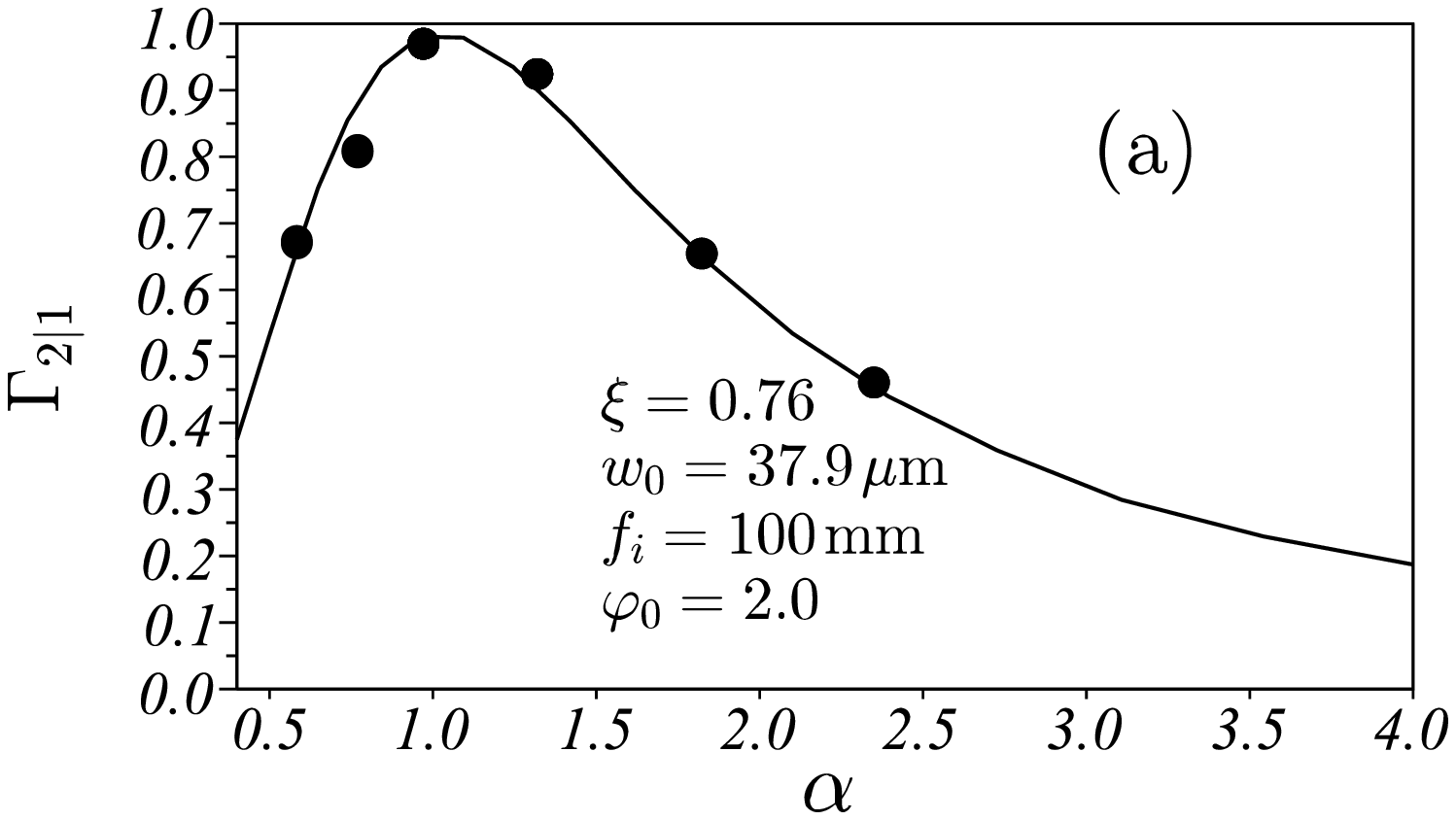}\\
	\includegraphics[width=\columnwidth]{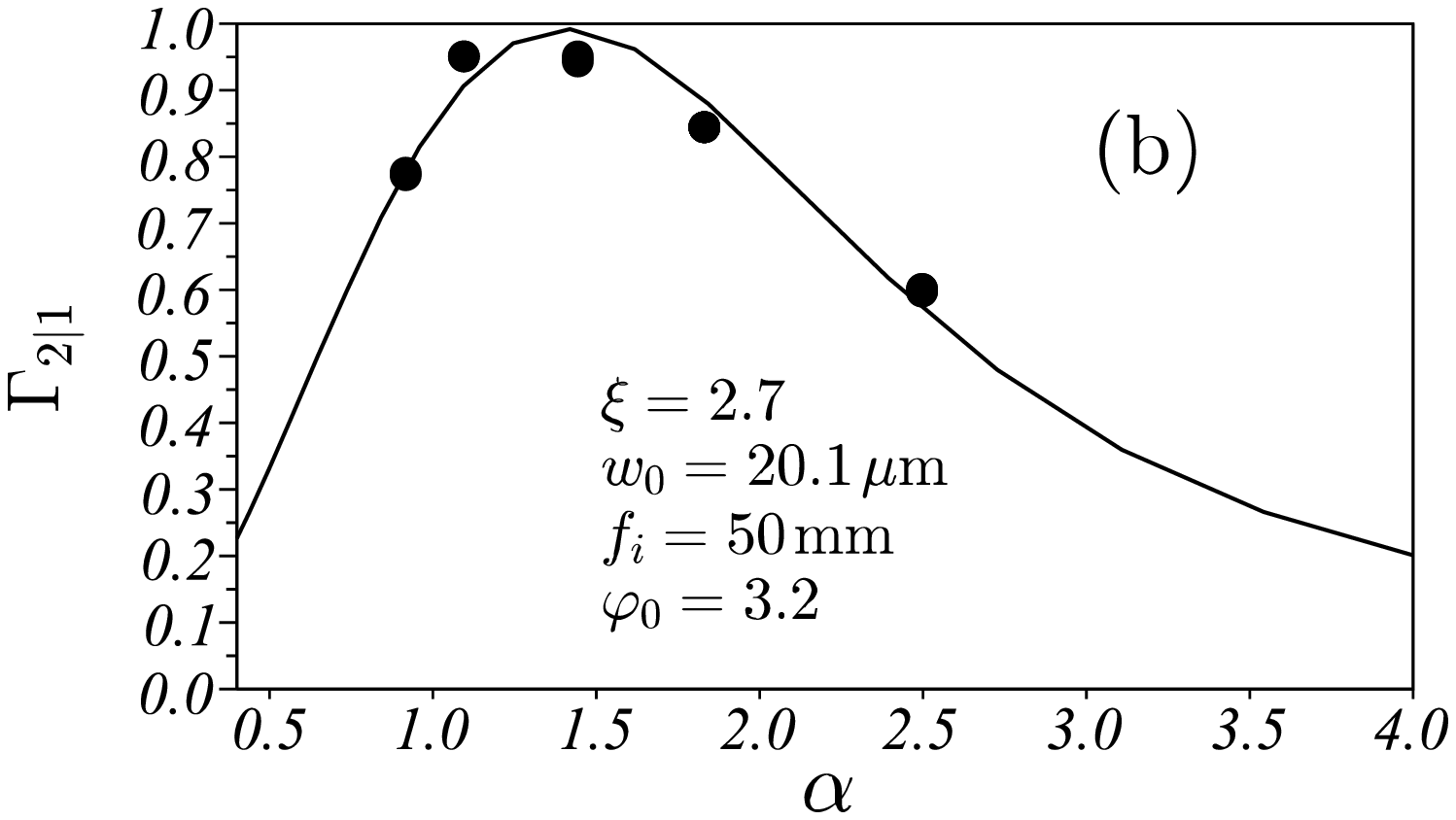}
	\caption{Comparison of theoretically (solid lines) vs. experimentally (dots) determined heralding ratio $\Gamma_{2|1}$, as a function of the normalized target mode waist $\vrho$ for two different focusing parameters~: $\veta = 0.76$ (a) and $\veta = 2.7$ (b).}
	\label{fig:ExperimentalResults1}
\end{figure}

The experimentally determined $\Gamma_{2|1}$ can then be compared to its theoretical value when the normalized target mode waist $\vrho$ is varied. Figure \ref{fig:ExperimentalResults1} shows two examples~: the pump focusing is kept constant with $f_p= 100$ mm ($\xi= 0.76$, Fig. \ref{fig:ExperimentalResults1}a) or $f_p=50$ mm ($\veta = 2.7$, Fig. \ref{fig:ExperimentalResults1}b). $\Gamma_{2|1}$ is normalized and plotted as a function of $\vrho$ for $\vzeta=0$ and the adequate value of the remaining unknown dimensionless parameter $\vphiz$ is found by horizontally fitting the theoretical curve to the experimental data. For $\veta = 0.76$, $\vphiz$ is found to be around $2.0$ while it is around $3.2$ for $\xi = 2.7$. These values of $\vphiz$ are also in agreement with the theoretical predictions corresponding to the brightness optimization. Indeed, although each plotted values of $\Gamma_{2|1}$ was obtained after optimizing this figure of merit itself, the experimental starting point was a preliminary optimization around a maximum source brightness. When $\Gamma_{2|1}$ and $P_2$ have a common optimum in the $(\vrho,\vphiz)$ space, it is not suprising to converge close to $(\vrho^\text{opt},\vphiz^\text{opt})$ (optimum of $P_2$ as shown in Fig. \ref{fig:TransmissionFactorSpatialOptimal} as a function of $\veta$) when optimizing with respect to $\Gamma_{2|1}$.

The very good agreement of our theory with the experimental results confirms its validity. For practical applications it is remarkable that only the measurement of the pump waist is required for the optimization of the SPDC source, the choice of the collection magnification being directly derived from the calculations presented in our analysis, which leaves the remaining parameter $\vphiz$ to a heuristic, using the mere crystal temperature.

\begin{figure}
	\centering
	\includegraphics[width=\columnwidth]{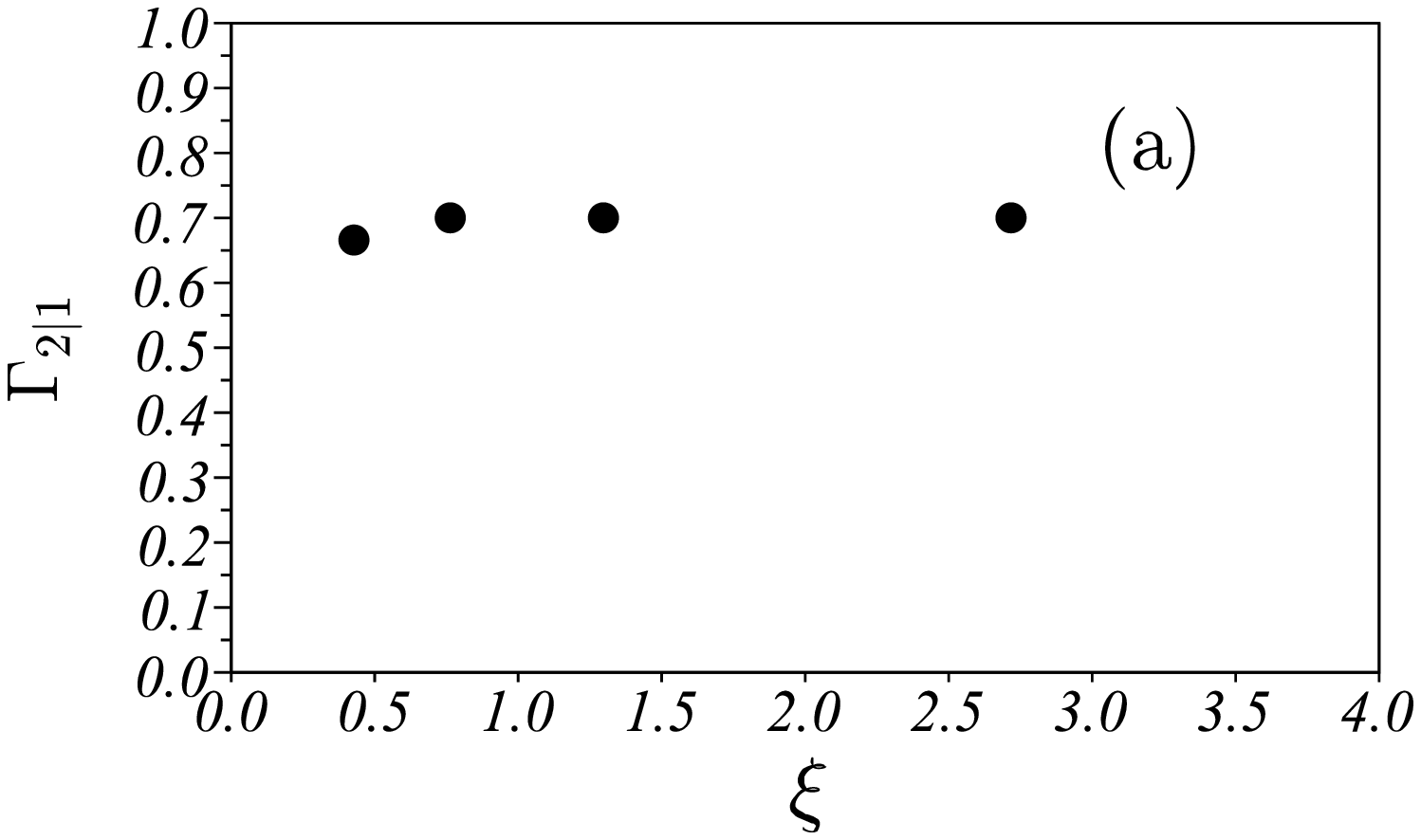}\\
	\includegraphics[width=\columnwidth]{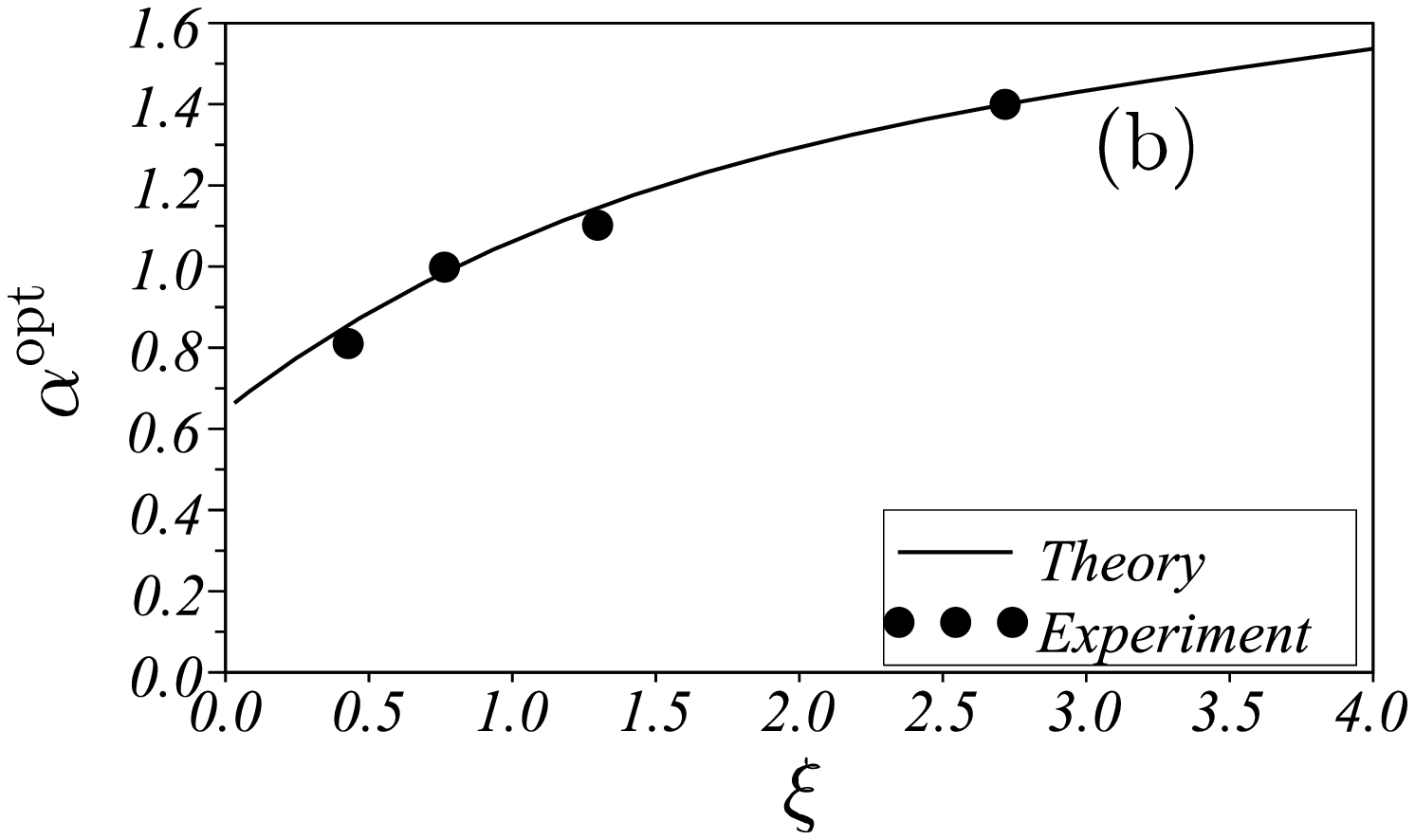}
	\caption{(a) Measured heralding ratio $\Gamma_{2|1}$ plotted for 4 focal lengths $f_p= 150, 100, 75, 50$ mm, as a function of the focusing parameter $\veta=0.44, 0.76, 1.05, 2.70$ . (b) Corresponding experimental (circles) and calculated (solid line) value of the normalized target mode waist $\vrho$ for each value of $\Gamma_{2|1}$}
\label{fig:ExperimentalResults2}
\end{figure}

The absolute value of the heralding ratio was also derived from the preliminary measurements of $i\ell_A=0.026$ and $i\ell_B=0.024$. In Fig. \ref{fig:ExperimentalResults2}a, the experimental absolute value of the heralding ratio $\Gamma_{2|1}$ is plotted as a function of the focusing parameter $\veta$. The result is almost constant, as predicted by the theory, but around $30$ \% below the expected optimal value that is close to $100$ \%. This difference is probably due to imperfections in the Gaussian pump beam and aberration in the optical system used to eliminate pump photons and to collect down-converted photon pairs into the fiber.

Let us also remark that the maximum measured value of $\Gamma_{2|1}= 0.7$, although not optimal, is to our knowledge impossible to reach by coupling PPLN waveguides into single-mode fibers, due to the mode mismatch between the rectangular profile of the guided mode and that -circular- of a fiber. This makes focused SPDC in bulk crystals probably more suitable than waveguide crystals when a high state fidelity is critical and limited by other constraints. For instance, the complex spectral filtering required for quantum memories associated with its limited storage-retrieval efficiency makes the coupling efficiency a parameter that is more than ever desirable to maximize.

In summary, we have confirmed 1) the dependence of the heralding ratio on the target mode size, 2) its quasi-independence on the focusing parameter, and 3) that our theory predicts with a very good precision the optimal target mode size for a particular focusing parameter.

\section{Discussion}\label{sec:Discussion}

\subsection{Comparison with other works}

As mentioned in the introduction, our work covers issues that were addressed by other authors with various assumptions or methods. The two figures of merit that can provide a relevant comparison are the absolute brightness and the coupling efficiency. We will also discuss a few reported experimental validations.

About the absolute brightness, Ling \emph{et al.} \cite{Ling2008} were among the first to propose an expression of the absolute photon pair rate collected into a single Gaussian spatial mode. Apart from the fact that they consider the case of a continuous pump ($\delta = 0$), our calculations are consistent dimension-wise. Moreover, everything else being equal, a spatial filtering factor can be identified in their analysis: its dependence with the focusing parameter and the longitudinal phase mismatch is of the form $\veta \sinc^2(\vphiz/2)$, which is indeed the result we find for the specific case of low focusing ($\veta \ll 1$) and $\vrho = \sqrt{2}$, corresponding to Ling's assumptions : thin crystal, negligible pump diffraction, equal Rayleigh lengths for pump and down-converted photons. However, under our less restrictive assumptions, we have shown that $\vrho = \sqrt{2}$ is optimal for $\veta = 2.84$, whereas for low focusing, $\vrho \leq 1$.

More recently, the absolute value of the source brightness was calculated by Mitchell \cite{Mitchell2009} in the Heisenberg picture, under assumptions similar to ours. This investigation has been restricted to a monochromatic, continuous pump field, and a direct expansion of the Hamiltonian onto the target Gaussian modes was performed, as opposed to our choice of investigating the single-mode coupling of free-space expanded fluorescence in the interaction picture. Our theory, which is developed in the Schrödinger picture, is nevertheless fully consistent with the results of Mitchell and we suggest in the next subsection how to generalize it using the same framework.

As for the coupling efficiency issue, let us recall that we investigated two parameters~: the pair collection efficiency and the heralding ratio. Our results concerning the pair collection efficiency are consistent with that of Bennink \cite{Bennink2010}, who found a linear dependence on the crystal length when the source bandwidth is much smaller than the phase-matching bandwidth, as opposed to Ljunggren \emph{et al.} \cite{Ljunggren2005} who conclude on a $\sqrt{L}$-dependence. Let us point out that, as Bennink \cite{Bennink2010}, we have taken into account the diffracting nature of the pump beam, and considered the longitudinal phase mismatch as a degree of freedom for optimization. As a result, our optimization of $\Gamma_2$ for Gaussian beams at the degenerate frequency is optimal close to the well-known Boyd \& Kleinman conditions \cite{Boyd1968}~: $\veta = 2.84$, $\vphiz = 3.2$ and equal Rayleigh length for the pump and down-converted photons ($\vrho = \sqrt{2}$). An advantage of our framework is that the particular pump and target modes, filter shape, pump linewidth, etc. are only used in the very last step of the optimization, \emph{i.e.} the computation of a mutliple integral. Up to then, our framework remains very general. 

As far as the heralding ratio is concerned, as opposed to the coupled brightness, it is found to be close to $1$ whatever the focusing strength, provided the target mode waist $\vrho$ and phase mismatch $\vphiz$ are adjusted according to our theory. Interestingly, a subset of $(\vrho,\vphiz)$ which maximizes the heralding ratio is generally close to the optimal brightness, giving the configuration for an optimal general source performance. This conclusion is different from that of Benninck \cite{Bennink2010} who finds that a strong reduction of brightness is necessary to achieve a high heralding ratio, highlighting what could be a fundamental trade-off for single-mode applications. This apparent disagreement could be due to our specific narrow-band assumption. Our experiments confirm that the narrow-band behaviour follows our predictions, which is a positive result as far as the perspectives of SPDC for narrow-band applications are concerned.

As remarked by Bennink, very few among the reported theoretical works have been experimentally validated, probably because of the complexity of such experiments that involve multiple parameters to optimize simultaneously and to measure precisely. Reported experiments were realized for a single set of non-optimal parameters \cite{Ljunggren2005} or in a configuration which is not in the scope of our theory \cite{Castelletto2005}. Thanks to a careful reduction of systematic errors, our experimental results show that our model could be used in order to optimize the experimental source configuration.

\subsection{Extension to other source designs}

Although the calculation done in Sec. \ref{sec:NarrowBand} uses a bulk periodically-poled crystal pumped with a single-mode diffractive gaussian pump beam, as required for successful comparison with our experiment, the general theoretical framework of Sec. \ref{sec:TheoreticalFramework} can be applied to other source designs. A few examples are given herafter.

First, it is straightforward to correct the results for non-degenerate down conversion, provided the frequency difference is small compared to the pump frequency.

It is also possible to use a non-gaussian pump or target mode profile, by changing the functions $\gespace$ and $\OO_0$ respectively. If required, a second dimensionless parameter can be introduced apart from $\vrho$ in order to characterize the mode ellipticity. In this case, the 4D- to 3D-integral reduction used in Eq. \eqref{eq:3Dintegral} is impossible due to the absence of cylindrical symmetry, making the computation longer. Let us note that for such high-dimensions integrals, a Monte-Carlo integration method could give faster results.

As far as the crystal type is concerned, as mentionned at the end of Sec. \ref{par:DownConvertedState}, a non-periodically or multi-periodically poled crystal can be used by coherently summing the first $N$ components of the Fourier series expansion of the non-linear susceptibility $\Chi(z)$. When $\Chi(z)$ is not of high complexity, the expansion can be truncated to a low $N$, making the calculation not much more time-consuming.

SPDC sources using a periodically-poled crystal with integrated waveguide are also commonly used. The coupling of photons down-converted in such devices could be modelled using our framework, by slicing the crystal into $N$ sub-crystals, for each of which the state coupled to the waveguide mode is calculated. For a slice $j=1 \cdots N$ centered on $z_j$, the target mode is assumed to have its waist located at $z_j$. If diffraction is taken into account as in the case of bulk crystals, the pump profile $\gespace(\rrho,z_j)$ differs in each slice according to Eq. \eqref{eq:Diffracting beam}. However, the pump field is generally also guided in the integrated waveguide and the pump profile should be chosen constant, which is done by removing the diffraction term in Eq. \eqref{eq:Diffracting beam}. 
The final state for which figures of merit like brightness or heralding ratio are evaluated results from coherently summing the states originating from each slice, taking into account their respective phases.

Let us remark that, in integrated waveguides designed to be single-mode for down-converted photons, the pump field is generally spatially multimode. This can also be taken into account by replacing the single-mode pump function $\gespace$ by a mode expansion $\sum_{j=1}^N {\gespace}_j$. As for a general nonlinear susceptibility distribution, this only increases the computation time by a factor $N$, which can be chosen according to the desired precision.

\section{Conclusion}

In this paper, we have calculated the state of a photon pair produced by a narrow-band spontaneous parametric down-conversion (SPDC) source with arbitrary pump spatial and temporal profile, and arbitrary filtering configuration.

When applied to Gaussian modes, our theory is consistent with the most recently reported work where realistic assumptions have been made. 
Within our assumptions, no incompatibility is observed between a high brightness and a high heralding ratio. We believe this result could increase the interest of SPDC sources for narrow-band quantum information applications.

Once validated under the Gaussian assumptions, our theoretical framework, which allows an extensive study of the source through many degrees of freedom, should allow the prediction of the best performance for more original source designs, including non-Gaussian pump and target modes, guided or diffractive, single or multi-mode, with arbitrary nonlinear susceptibility longitudinal distribution.

\begin{acknowledgments}
This work is a part of the project ``embryonic Quantum Network'', funded by the ``Agence Nationale pour la Recherche''.
\end{acknowledgments}

\appendix

\section{Conventions for Fourier transforms and convolutions}\label{ann:FourierConventions}

All along this paper, we use the following Fourier transform conventions~:
\begin{align*}
	\tf{f}(\w) =\intt e^{\j \w t} f(t) &\longleftrightarrow f(t)=\intw e^{-\j \w t} \tf{f}(\w)\\
	\tf{f}(\kk)=\intr e^{-\j \kk \cdot \rr}f(\rr) &\longleftrightarrow f(\rr)=\intk e^{\j \kk \cdot \rr} \tf{f}(\kk)
\end{align*}

$f$ is normalized if~:
\begin{align*}
	\intt |f(t)|^2 &= \intw |\tf{f}(\w)|^2 = 1 \\
	\intr |f(\rr)|^2 &= \intk |\tf{f}(\kk)|^2 = 1
\end{align*}

Convolutions are defined as follows~:
\begin{align*}
	f \! * \! g(t) &= \int\!dt'\, f(t')g(t-t') \\
	f \! * \! g(\rr) &= \int\!d^3\rr'\, f(\rr')g(\rr-\rr') \\
	\tf{f} \! * \! \tf{g}(\w) &= \int\!\frac{d\w'}{2\pi}\, \tf{f}(\w') \tf{g}(\w-\w') \\
	\tf{f} \! * \! \tf{g}(\kk) &= \int\!\frac{d^3\kk'}{(2\pi)^3}\, \tf{f}(\kk') \tf{g}(\kk-\kk')
\end{align*}

\section{Spatial and spectral filtering}\label{ann:OpticalElements}

In this appendix we describe the effect of filters on a single photon state. Generalization to the case of photon pairs is made in Section \ref{par:Filtering}.

Although the action of filters will be described in the Fourier space, they are located at a particular position of the setup and a specific time lag $[t_0,t_1]$ can be defined such that when $t < t_0$, the photon state is completely unaffected by the filter yet and when $t > t_1$, the effect of the filter on the photon state has been completed. After having introduced some tools in the first two sections, we will successively examine how spectral and spatial filters change a state $\ket{\Psi(t_0)}$ into a state $\ket{\Psi(t_1)}$.

\subsection{``Localized'' photonic states and pseudo-wavefunction}\label{ann:LocalizedStates}

In this paper, we sometimes use localized photonic states introduced by L. Mandel \cite{Mandel1966}. Based on plane-wave states $\ket{1_\kk}$ in a quantization volume $\volq$, a state describing a photon localized around $\rr$ is defined by~:
\begin{equation}\label{eq:fluo:EtatLocaliseDefinition}
	\ket{1_{\rr}} = \sum_\kk \frac{e^{- \j \kk \cdot \rr}}{\sqrt{\volq}} \ket{1_{\kk}}
\end{equation}

Then, a general photon state $\ket{\psi} = \sum_\kk \psi_\kk \ket{1_{\kk}}$ can be written as
\begin{equation}
	\ket{\psi} = \intr[\volq] \psi(\rr) \ket{1_{\rr}}
\end{equation}
where $\psi(\rr)$ is a spatial pseudo-wavefunction for which $\psi_\kk$ are the coefficients of its Fourier series expansion in the quantization volume $\volq$. Such a wavefunction is only valid provided the volume $\vol$ in which the localization probability $P(\vol)$, defined as follows, is evaluated is large enough (each dimension much larger that the wavelength)~:
\begin{equation}
	P(\vol) = \intr[\vol] \abs{ \braket{1_{\rr}}{\psi} }^2 = \intr[\vol] \abs{ \psi(\rr) }^2
\end{equation}

This probability is naturally equal to unity in the quantization volume~:
\begin{equation}
	P(\volq) = \intr[\volq] \abs{ \psi(\rr) }^2 = \braket{\psi}{\psi} = \sum_\kk \abs{ \psi_\kk }^2 = 1.
\end{equation}

\subsection{Spatial filtering}

\subsubsection{Principle}

A spatial filter located on the propagation axis at $z=z_0$ is modelled as the coupling in the plane $z=z_0$ of the down-converted field into a single spatial mode defined by a function $\OO_{\w,0}(\rr)$. The index $\w$ indicates that this spatial mode can be frequency-dependent for a given filter.

Note that $z_0$ may be in $[-L/2,L/2]$. As an example, if the spatial filtering is done via an optical fiber, the effective location of the filter is where the lens collection system images the entrance of the fiber.

To evaluate its transmitted component, the down-converted field initially described as a superposition of plane waves is better expanded on a particular set of orthogonal modes $\{ \OO_{\w,j}(\rr) \}$, one of which ($\OO_{\w,0}(\rr)$) being the spatial mode selected by the considered filter. 

If coupled to other modes, photons are supposed to be lost. As for spectral filtering, this leads to a mixed state.

For instance, the fundamental mode can be the Gaussian mode of the Laguerre-Gauss basis, suited to single-mode fibers. Then $\OO_{\w,j=0}(\rr)$ describes a Gaussian beam of waist size equal to the field radius of a fiber mode. If the image of the fiber entrance is located at $z=z_0$, $\OO_{\w,j=0}(\rrho,z_0)$ describes the Gaussian beam waist profile corresponding to the fiber transverse mode.

A projection at $z=z_0$ into that single mode selects the photon state component which is transmitted by the filter. Once transmitted, the state is said to be in the spatial mode characterized by the annihilation operator $\aao[\w]$. Let us note that whether the spatial mode for $z \geq z_0$ continues to be that of a Gaussian beam (as for a cavity) or is actually described by a guided propagation with a constant transverse profile $\OO_{\w,0}(\rrho,z_0)$ and propagation constant $\beta_\w = n_f(\w) \w/c$ like in a single mode fiber of effective refractive index $\nf$ \cite{Ghatak1998} will not change the description:
\begin{equation}\label{eq:OO}
	\OO_{\w,j=0}(\rrho, z \geq z_0) = \OO_{\w,0}(\rrho,z_0) e^{\j \beta_\w (z-z_0)}
\end{equation}

In a quantization volume $\volq = \SS \times \LL$, a field in the mode $\OO_{\w,j}(\rr)$ will be said to be in the following quantum state~:
\begin{equation}
	\ket{\OO_{\w,j}} = \frac{1}{\sqrt{\LL}} \intr \OO_{\w,j}(\rr) \ket{1_{\rr}}
\end{equation}
so that the states are normalized~: $\braket{\OO_{\w,j}}{\OO_{\w,j'}} = \dirac_{jj'}$. Functions $\OO_{\w,j}(\rr)$ are only transversally normalized~: $\int d^2\rrho \OO^*_{\w,j'}(\rrho,z) \OO_{\w,j}(\rrho,z) = \dirac_{jj'}$ and $\int_{\volq} d^3\rr \OO^*_{\w,j'}(\rr) \OO_{\w,j}(\rr) = \LL \, \dirac_{jj'}$.

\subsubsection{Calculation of the transmitted field component }

In a quantization volume containing a crystal before $z=z_0$ and a spatial filter at $z_0$, a one-photon field in mode $\ell$ is in the state~:
\begin{align}
	\ket{1_{\ell}} &= \intr f(\rr) \ket{1_{\rr}} \\
		&= \intr \Big(\Theta(z_0 \! - \! z) \frac{e^{\j \kk_{\ell} \cdot \rr}} {\sqrt{\volq}} \notag \\
		&+ \Theta(z \! - \! z_0) \sum_{j} g_j(\ell) \frac{\OO_{\w_{\ell},j}(\rr)}{\sqrt{\LL}} \Big) \ket{1_{\rr}}
\end{align}
where $\Theta(z)$ is Heaviside's function and $g_{j}(\ell)$ are determined by the boundary conditions at $z=z_0$~: 
\begin{equation}
	\sum_{j} g_j(\ell) \frac{\OO_{\w_{\ell},j}(\rrho,z_0)}{\sqrt{\LL}} = \frac{e^{\j \kkappa'_{\ell}\cdot\rrho} e^{\j k'_{z,\ell}\cdot z_0}}{\sqrt{\volq}}
\end{equation}
Remark that the wavevector takes into account the change of medium according to Snell-Descartes' law of refraction~:
\begin{subequations}
	\begin{equation}
		\frac{\kkappa'}{\kkappa} = \frac{|\kk'|\sin \theta'}{|\kk| \sin \theta} = 1 \qquad
		\frac{k'_z}{k_z} = \frac{|\kk'| \cos \theta'}{|\kk|  \cos \theta} \approx \frac{\nf}{\nc} \label{eq:SnellDescartes2}
	\end{equation}
\end{subequations}

where $\nf$ is the refractive index of the medium in which $\OO_{\w_{\ell},j}(\rr)$ describes the spatial mode and $\nc$ the refractive index in the crystal.

Using the orthonormality of functions $\{ \OO_{\w,j}(\rr) \}$, one gets~:
\begin{align}
	g_j(\ell) &= \int d^2 \rrho \OO^*_{\w_{\ell},j}(\rrho,z_0) \frac{e^{\j \kkappa_{\ell}\cdot\rrho} e^{\j k'_{z,\ell}\cdot z_0}}{\sqrt{\volq}} \sqrt{\LL} \notag \\
	&= \frac{1}{\sqrt{\SS}} e^{\j k'_{z,\ell} \cdot z_0} \TOO^*_{\w_{\ell},j}(\kkappa_\ell,z_0)
\end{align}
where $\TOO_{\w_{\ell},j}(\kkappa,z_0)$ is the Fourier transform of $\OO_{\w_{\ell},j}(\rrho,z_0)$ and $^*$ designates a complex conjugate.

Starting from an initial state $\ket{\Psi(t_0)} = \sum_{\ell} \mu_\ell \ket{1_\ell}$ localized at $z<z_0$, the state becomes for $z>z_0$~:
\begin{equation}
	\ket{\Psi(t_1)} = \sum_{j} \sum_{\ell} \mu_\ell e^{\j k'_{z,\ell} \cdot z_0} \frac{1}{\sqrt{\SS}} \TOO^*_{\w_{\ell},j} (\kkappa_{\ell},z_0) \ket{\OO_{\w_{\ell},j}}
\end{equation}
Ignoring components of the state which are not within the mode $\ket{\OO_{\w_{\ell},0}}$ transmitted by the filter, the transmitted one-photon state becomes~:
\begin{multline}
	\ket{\Psi(t_1)} = \sum_{\ell} \mu_\ell e^{\j k'_{z,\ell} \cdot z_0} \frac{1}{\sqrt{\SS}} \TOO^*_{\w_{\ell},0} (\kkappa_{\ell},z_0) \ket{\OO_{\w_{\ell},0}} \\
		+ \text{\emph{ignored zero-photon states}}
\end{multline}

Note that if one of the functions $\{ \OO_{\w,j}(\rr) \}$ describes the spatial eigenmode of any other filter (\emph{e.g.} a rectangular waveguide), the result is still valid.

As for spectral filtering, extrapolation to two-photon states is straightforward.


%



\begin{thebibliography}{51}%
\makeatletter
\providecommand \@ifxundefined [1]{%
 \@ifx{#1\undefined}
}%
\providecommand \@ifnum [1]{%
 \ifnum #1\expandafter \@firstoftwo
 \else \expandafter \@secondoftwo
 \fi
}%
\providecommand \@ifx [1]{%
 \ifx #1\expandafter \@firstoftwo
 \else \expandafter \@secondoftwo
 \fi
}%
\providecommand \natexlab [1]{#1}%
\providecommand \enquote  [1]{``#1''}%
\providecommand \bibnamefont  [1]{#1}%
\providecommand \bibfnamefont [1]{#1}%
\providecommand \citenamefont [1]{#1}%
\providecommand \href@noop [0]{\@secondoftwo}%
\providecommand \href [0]{\begingroup \@sanitize@url \@href}%
\providecommand \@href[1]{\@@startlink{#1}\@@href}%
\providecommand \@@href[1]{\endgroup#1\@@endlink}%
\providecommand \@sanitize@url [0]{\catcode `\\12\catcode `\$12\catcode
  `\&12\catcode `\#12\catcode `\^12\catcode `\_12\catcode `\%12\relax}%
\providecommand \@@startlink[1]{}%
\providecommand \@@endlink[0]{}%
\providecommand \url  [0]{\begingroup\@sanitize@url \@url }%
\providecommand \@url [1]{\endgroup\@href {#1}{\urlprefix }}%
\providecommand \urlprefix  [0]{URL }%
\providecommand \Eprint [0]{\href }%
\providecommand \doibase [0]{http://dx.doi.org/}%
\providecommand \selectlanguage [0]{\@gobble}%
\providecommand \bibinfo  [0]{\@secondoftwo}%
\providecommand \bibfield  [0]{\@secondoftwo}%
\providecommand \translation [1]{[#1]}%
\providecommand \BibitemOpen [0]{}%
\providecommand \bibitemStop [0]{}%
\providecommand \bibitemNoStop [0]{.\EOS\space}%
\providecommand \EOS [0]{\spacefactor3000\relax}%
\providecommand \BibitemShut  [1]{\csname bibitem#1\endcsname}%
\let\auto@bib@innerbib\@empty
\bibitem [{\citenamefont {Tittel}\ and\ \citenamefont
  {Weihs}(2001)}]{Tittel2001}%
  \BibitemOpen
  \bibfield  {author} {\bibinfo {author} {\bibfnamefont {W.}~\bibnamefont {Tittel}}\  and\  
\bibinfo {author} {\bibfnamefont {G.}~\bibnamefont {Weihs}},\ }\href@noop {} 
{\bibfield  {journal} {\bibinfo  {journal} {Quantum Information and Computation}\ }
\textbf {\bibinfo {volume} {1}},\ \bibinfo {pages} {3}\ (\bibinfo {year} {2001})}\BibitemShut {NoStop}%
\bibitem [{\citenamefont {Gr{\"o}blacher}\ \emph {et~al.}(2007)\citenamefont
  {Gr{\"o}blacher}, \citenamefont {Paterek}, \citenamefont {Kaltenbaek} \emph
  {et~al.}}]{Groblacher2007}%
  \BibitemOpen
  \bibfield  {author} {\bibinfo {author} {\bibfnamefont {S.}~\bibnamefont
  {Gr{\"o}blacher}}, \bibinfo {author} {\bibfnamefont {T.}~\bibnamefont
  {Paterek}}, \bibinfo {author} {\bibfnamefont {R.}~\bibnamefont {Kaltenbaek}},
   \emph {et~al.},\ }\href@noop {} {\bibfield  {journal} {\bibinfo  {journal}
  {Nature}\ }\textbf {\bibinfo {volume} {446}},\ \bibinfo {pages} {871}
  (\bibinfo {year} {2007})}\BibitemShut {NoStop}%
\bibitem [{\citenamefont {Gisin}\ and\ \citenamefont {Thew}(2007)}]{Gisin2007}%
 \BibitemOpen
  \bibfield  {author} {\bibinfo {author} {\bibfnamefont {N.}~\bibnamefont
  {Gisin}}\ and\ \bibinfo {author} {\bibfnamefont {R.}~\bibnamefont {Thew}},\
  }
{\bibfield  {journal}
  {\bibinfo  {journal} {Nature Photonics}\ }\textbf {\bibinfo {volume} {1}},\
  \bibinfo {pages} {165} (\bibinfo {year} {2007})},\ \bibinfo {note}
  {10.1038/nphoton.2007.22}\BibitemShut {NoStop}%
\bibitem [{\citenamefont {Gisin}\ and\ \citenamefont {Thew}(2010)}]{Gisin2010}%
  \BibitemOpen
  \bibfield  {author} {\bibinfo {author} {\bibfnamefont {N.}~\bibnamefont
  {Gisin}}\ and\ \bibinfo {author} {\bibfnamefont {R.~T.}\ \bibnamefont
  {Thew}},\ }
{\bibfield  {journal}
  {\bibinfo  {journal} {Electronics Letters}\ }\textbf {\bibinfo {volume}
  {46}},\ \bibinfo {pages} {965} (\bibinfo {year} {2010})}\BibitemShut
  {NoStop}%
\bibitem [{\citenamefont {Ekert}(1991)}]{Ekert1991}%
  \BibitemOpen
  \bibfield  {author} {\bibinfo {author} {\bibfnamefont {A.~K.}\ \bibnamefont
  {Ekert}},\ }
{\bibfield  {journal}
  {\bibinfo  {journal} {Phys. Rev. Lett.}\ }\textbf {\bibinfo {volume} {67}},\
  \bibinfo {pages} {661} (\bibinfo {year} {1991})}\BibitemShut {NoStop}%
\bibitem [{\citenamefont {Shapiro}(2002)}]{Shapiro2002}%
  \BibitemOpen
  \bibfield  {author} {\bibinfo {author} {\bibfnamefont {J.~H.}\ \bibnamefont
  {Shapiro}},\ }
{\bibfield
  {journal} {\bibinfo  {journal} {New Journal of Physics}\ }\textbf {\bibinfo
  {volume} {4}},\ \bibinfo {pages} {47} (\bibinfo {year} {2002})}\BibitemShut
  {NoStop}%
\bibitem [{\citenamefont {Simon}\ \emph {et~al.}(2007)\citenamefont {Simon},
  \citenamefont {de~Riedmatten}, \citenamefont {Afzelius}, \citenamefont
  {Sangouard}, \citenamefont {Zbinden},\ and\ \citenamefont
  {Gisin}}]{Simon2007}%
  \BibitemOpen
  \bibfield  {author} {\bibinfo {author} {\bibfnamefont {C.}~\bibnamefont
  {Simon}}, \bibinfo {author} {\bibfnamefont {H.}~\bibnamefont
  {de~Riedmatten}}, \bibinfo {author} {\bibfnamefont {M.}~\bibnamefont
  {Afzelius}}, \bibinfo {author} {\bibfnamefont {N.}~\bibnamefont {Sangouard}},
  \bibinfo {author} {\bibfnamefont {H.}~\bibnamefont {Zbinden}}, \ and\
  \bibinfo {author} {\bibfnamefont {N.}~\bibnamefont {Gisin}},\ }\href@noop {}
  {\bibfield  {journal} {\bibinfo  {journal} {Physical Review Letters} \ }\textbf {\bibinfo
  {volume} {98}},\ \bibinfo {pages} {190503}
  (\bibinfo {year} {2007})}\BibitemShut {NoStop}%
\bibitem [{\citenamefont {Wong}\ \emph {et~al.}(2006)\citenamefont {Wong},
  \citenamefont {Shapiro},\ and\ \citenamefont {Kim}}]{Wong2006}%
  \BibitemOpen
  \bibfield  {author} {\bibinfo {author} {\bibfnamefont {F.}~\bibnamefont
  {Wong}}, \bibinfo {author} {\bibfnamefont {J.}~\bibnamefont {Shapiro}}, \
  and\ \bibinfo {author} {\bibfnamefont {T.}~\bibnamefont {Kim}},\ }
{\bibfield  {journal} {\bibinfo
  {journal} {Laser Physics}\ }\textbf {\bibinfo {volume} {16}},\ \bibinfo
  {pages} {1517} (\bibinfo {year} {2006})}\BibitemShut {NoStop}%
\bibitem [{\citenamefont {Kwiat}\ \emph {et~al.}(1995)\citenamefont {Kwiat},
  \citenamefont {Mattle}, \citenamefont {Weinfurter}, \citenamefont
  {Zeilinger}, \citenamefont {Sergienko},\ and\ \citenamefont
  {Shih}}]{Kwiat1995}%
  \BibitemOpen
  \bibfield  {author} {\bibinfo {author} {\bibfnamefont {P.~G.}\ \bibnamefont
  {Kwiat}}, \bibinfo {author} {\bibfnamefont {K.}~\bibnamefont {Mattle}},
  \bibinfo {author} {\bibfnamefont {H.}~\bibnamefont {Weinfurter}}, \bibinfo
  {author} {\bibfnamefont {A.}~\bibnamefont {Zeilinger}}, \bibinfo {author}
  {\bibfnamefont {A.~V.}\ \bibnamefont {Sergienko}}, \ and\ \bibinfo {author}
  {\bibfnamefont {Y.}~\bibnamefont {Shih}},\ }
{\bibfield  {journal} {\bibinfo  {journal}
  {Phys. Rev. Lett.}\ }\textbf {\bibinfo {volume} {75}},\ \bibinfo {pages}
  {4337} (\bibinfo {year} {1995})}\BibitemShut {NoStop}%
\bibitem [{\citenamefont {Noh}\ \emph {et~al.}(2007)\citenamefont {Noh},
  \citenamefont {Kim}, \citenamefont {Zyung},\ and\ \citenamefont
  {Kim}}]{Noh2007}%
  \BibitemOpen
  \bibfield  {author} {\bibinfo {author} {\bibfnamefont {T.~G.}\ \bibnamefont
  {Noh}}, \bibinfo {author} {\bibfnamefont {H.}~\bibnamefont {Kim}}, \bibinfo
  {author} {\bibfnamefont {T.}~\bibnamefont {Zyung}}, \ and\ \bibinfo {author}
  {\bibfnamefont {J.}~\bibnamefont {Kim}},\ }\href@noop {} {\bibfield
  {journal} {\bibinfo  {journal} {Appl. Phys. Lett.}\ }\textbf {\bibinfo
  {volume} {90}},\ \bibinfo {pages} {011116} (\bibinfo {year}
  {2007})}\BibitemShut {NoStop}%
\bibitem [{\citenamefont {Altepeter}\ \emph {et~al.}(2005)\citenamefont
  {Altepeter}, \citenamefont {Jeffrey},\ and\ \citenamefont
  {Kwiat}}]{Altepeter2005}%
  \BibitemOpen
  \bibfield  {author} {\bibinfo {author} {\bibfnamefont {J.}~\bibnamefont
  {Altepeter}}, \bibinfo {author} {\bibfnamefont {E.}~\bibnamefont {Jeffrey}},
  \ and\ \bibinfo {author} {\bibfnamefont {P.}~\bibnamefont {Kwiat}},\
  }\href@noop {} {\bibfield  {journal} {\bibinfo  {journal} {Opt. Express}\
  }\textbf {\bibinfo {volume} {13}},\ \bibinfo {pages} {8951} (\bibinfo {year}
  {2005})}\BibitemShut {NoStop}%
\bibitem [{\citenamefont {Shi}\ and\ \citenamefont {Tomita}(2004)}]{Shi2004}%
  \BibitemOpen
  \bibfield  {author} {\bibinfo {author} {\bibfnamefont {B.}~\bibnamefont
  {Shi}}\ and\ \bibinfo {author} {\bibfnamefont {A.}~\bibnamefont {Tomita}},\
  }\href@noop {} {\bibfield  {journal} {\bibinfo  {journal} {J. Opt. Soc. Am.
  B}\ }\textbf {\bibinfo {volume} {21}},\ \bibinfo {pages} {2081} (\bibinfo
  {year} {2004})}\BibitemShut {NoStop}%
\bibitem [{\citenamefont {Guillet~de Chatellus}\ \emph
  {et~al.}(2006)\citenamefont {Guillet~de Chatellus}, \citenamefont
  {Sergienko}, \citenamefont {Saleh}, \citenamefont {Teich},\ and\
  \citenamefont {Di~Giuseppe}}]{Guillet2006}%
  \BibitemOpen
  \bibfield  {author} {\bibinfo {author} {\bibfnamefont {H.}~\bibnamefont
  {Guillet~de Chatellus}}, \bibinfo {author} {\bibfnamefont {A.}~\bibnamefont
  {Sergienko}}, \bibinfo {author} {\bibfnamefont {B.}~\bibnamefont {Saleh}},
  \bibinfo {author} {\bibfnamefont {M.}~\bibnamefont {Teich}}, \ and\ \bibinfo
  {author} {\bibfnamefont {G.}~\bibnamefont {Di~Giuseppe}},\ }\href@noop {}
  {\bibfield  {journal} {\bibinfo  {journal} {Opt. Express}\ }\textbf {\bibinfo
  {volume} {14}},\ \bibinfo {pages} {10060} (\bibinfo {year}
  {2006})}\BibitemShut {NoStop}%
\bibitem [{\citenamefont {Fiorentino}\ \emph {et~al.}(2005)\citenamefont
  {Fiorentino}, \citenamefont {Kuklewicz},\ and\ \citenamefont
  {Wong}}]{Fiorentino2005}%
  \BibitemOpen
  \bibfield  {author} {\bibinfo {author} {\bibfnamefont {M.}~\bibnamefont
  {Fiorentino}}, \bibinfo {author} {\bibfnamefont {C.}~\bibnamefont
  {Kuklewicz}}, \ and\ \bibinfo {author} {\bibfnamefont {F.}~\bibnamefont
  {Wong}},\ }\href@noop {} {\bibfield  {journal} {\bibinfo  {journal} {Opt.
  Express}\ }\textbf {\bibinfo {volume} {13}},\ \bibinfo {pages} {127}
  (\bibinfo {year} {2005})}\BibitemShut {NoStop}%
\bibitem [{\citenamefont {Fedrizzi}\ \emph {et~al.}(2007)\citenamefont
  {Fedrizzi}, \citenamefont {Herbst}, \citenamefont {Poppe}, \citenamefont
  {Jennewein},\ and\ \citenamefont {Zeilinger}}]{Fedrizzi2007}%
  \BibitemOpen
  \bibfield  {author} {\bibinfo {author} {\bibfnamefont {A.}~\bibnamefont
  {Fedrizzi}}, \bibinfo {author} {\bibfnamefont {T.}~\bibnamefont {Herbst}},
  \bibinfo {author} {\bibfnamefont {A.}~\bibnamefont {Poppe}}, \bibinfo
  {author} {\bibfnamefont {T.}~\bibnamefont {Jennewein}}, \ and\ \bibinfo
  {author} {\bibfnamefont {A.}~\bibnamefont {Zeilinger}},\ }
 {\bibfield{journal} {\bibinfo  {journal} {Opt. Express}\ }\textbf {\bibinfo {volume}
  {15}},\ \bibinfo {pages} {15377} (\bibinfo {year} {2007})}\BibitemShut
  {NoStop}%
\bibitem [{\citenamefont {Kuzucu}\ and\ \citenamefont
  {Wong}(2008)}]{Kuzucu2008}%
  \BibitemOpen
  \bibfield  {author} {\bibinfo {author} {\bibfnamefont {O.}~\bibnamefont
  {Kuzucu}}\ and\ \bibinfo {author} {\bibfnamefont {F.~N.~C.}\ \bibnamefont
  {Wong}},\ }
{\bibfield  {journal}
  {\bibinfo  {journal} {Phys. Rev. A}\ }\textbf {\bibinfo {volume} {77}},\
  \bibinfo {pages} {032314} (\bibinfo {year} {2008})}\BibitemShut {NoStop}%
\bibitem [{\citenamefont {Virally}\ \emph {et~al.}(2010)\citenamefont
  {Virally}, \citenamefont {Lacroix},\ and\ \citenamefont
  {Godbout}}]{Virally2010}%
  \BibitemOpen
  \bibfield  {author} {\bibinfo {author} {\bibfnamefont {S.}~\bibnamefont
  {Virally}}, \bibinfo {author} {\bibfnamefont {S.}~\bibnamefont {Lacroix}}, \
  and\ \bibinfo {author} {\bibfnamefont {N.}~\bibnamefont {Godbout}},\ }
{\bibfield  {journal} {\bibinfo
  {journal} {Phys. Rev. A}\ }\textbf {\bibinfo {volume} {81}},\ \bibinfo
  {pages} {013808} (\bibinfo {year} {2010})}\BibitemShut {NoStop}%
\bibitem [{\citenamefont {Smirr}\ \emph
  {et~al.}(2011{\natexlab{a}})\citenamefont {Smirr}, \citenamefont {Frey},
  \citenamefont {Diamanti}, \citenamefont {All\'{e}aume},\ and\ \citenamefont
  {Zaquine}}]{Smirr2010a}%
  \BibitemOpen
  \bibfield  {author} {\bibinfo {author} {\bibfnamefont {J.~L.}\ \bibnamefont
  {Smirr}}, \bibinfo {author} {\bibfnamefont {R.}~\bibnamefont {Frey}},
  \bibinfo {author} {\bibfnamefont {E.}~\bibnamefont {Diamanti}}, \bibinfo
  {author} {\bibfnamefont {R.}~\bibnamefont {All\'{e}aume}}, \ and\ \bibinfo
  {author} {\bibfnamefont {I.}~\bibnamefont {Zaquine}},\ }
{\bibfield  {journal} {\bibinfo  {journal} {J. Opt.
  Soc. Am. B}\ }\textbf {\bibinfo {volume} {28}},\ \bibinfo {pages} {832}
  (\bibinfo {year} {2011}{\natexlab{a}})}\BibitemShut {NoStop}%
\bibitem [{\citenamefont {Lvovsky}\ \emph {et~al.}(2009)\citenamefont
  {Lvovsky}, \citenamefont {Sanders},\ and\ \citenamefont
  {Tittel}}]{Lvovsky2009}%
  \BibitemOpen
  \bibfield  {author} {\bibinfo {author} {\bibfnamefont {A.}~\bibnamefont
  {Lvovsky}}, \bibinfo {author} {\bibfnamefont {B.}~\bibnamefont {Sanders}}, \
  and\ \bibinfo {author} {\bibfnamefont {W.}~\bibnamefont {Tittel}},\
  }\href@noop {} {\bibfield  {journal} {\bibinfo  {journal} {Nature Photonics}\
  }\textbf {\bibinfo {volume} {3}},\ \bibinfo {pages} {706} (\bibinfo {year}
  {2009})}\BibitemShut {NoStop}%
\bibitem [{\citenamefont {{Simon, C.}}\ \emph {et~al.}(2010)\citenamefont
  {{Simon, C.}}, \citenamefont {{Afzelius, M.}}, \citenamefont {{Appel, J.}},
  \citenamefont {{Boyer de la Giroday, A.}}, \citenamefont {{Dewhurst, S. J.}},
  \citenamefont {{Gisin, N.}}, \citenamefont {{Hu, C. Y.}}, \citenamefont
  {{Jelezko, F.}}, \citenamefont {{Kr\"oll, S.}}, \citenamefont {{M\"uller, J.
  H.}}, \citenamefont {{Nunn, J.}}, \citenamefont {{Polzik, E. S.}},
  \citenamefont {{Rarity, J. G.}}, \citenamefont {{De Riedmatten, H.}},
  \citenamefont {{Rosenfeld, W.}}, \citenamefont {{Shields, A. J.}},
  \citenamefont {{Sk\"old, N.}}, \citenamefont {{Stevenson, R. M.}},
  \citenamefont {{Thew, R.}}, \citenamefont {{Walmsley, I. A.}}, \citenamefont
  {{Weber, M. C.}}, \citenamefont {{Weinfurter, H.}}, \citenamefont
  {{Wrachtrup, J.}},\ and\ \citenamefont {{Young, R. J.}}}]{Simon2010}%
  \BibitemOpen
  \bibfield  {author} {\bibinfo {author} {\bibnamefont {{Simon, C.}}}, \bibinfo
  {author} {\bibnamefont {{Afzelius, M.}}}, \bibinfo {author} {\bibnamefont
  {{Appel, J.}}}, \bibinfo {author} {\bibnamefont {{Boyer de la Giroday, A.}}},
  \bibinfo {author} {\bibnamefont {{Dewhurst, S. J.}}}, \bibinfo {author}
  {\bibnamefont {{Gisin, N.}}}, \bibinfo {author} {\bibnamefont {{Hu, C. Y.}}},
  \bibinfo {author} {\bibnamefont {{Jelezko, F.}}}, \bibinfo {author}
  {\bibnamefont {{Kr\"oll, S.}}}, \bibinfo {author} {\bibnamefont {{M\"uller,
  J. H.}}}, \bibinfo {author} {\bibnamefont {{Nunn, J.}}}, \bibinfo {author}
  {\bibnamefont {{Polzik, E. S.}}}, \bibinfo {author} {\bibnamefont {{Rarity,
  J. G.}}}, \bibinfo {author} {\bibnamefont {{De Riedmatten, H.}}}, \bibinfo
  {author} {\bibnamefont {{Rosenfeld, W.}}}, \bibinfo {author} {\bibnamefont
  {{Shields, A. J.}}}, \bibinfo {author} {\bibnamefont {{Sk\"old, N.}}},
  \bibinfo {author} {\bibnamefont {{Stevenson, R. M.}}}, \bibinfo {author}
  {\bibnamefont {{Thew, R.}}}, \bibinfo {author} {\bibnamefont {{Walmsley, I.
  A.}}}, \bibinfo {author} {\bibnamefont {{Weber, M. C.}}}, \bibinfo {author}
  {\bibnamefont {{Weinfurter, H.}}}, \bibinfo {author} {\bibnamefont
  {{Wrachtrup, J.}}}, \ and\ \bibinfo {author} {\bibnamefont {{Young, R.
  J.}}},\ }
{\bibfield  {journal}
  {\bibinfo  {journal} {Eur. Phys. J. D}\ }\textbf {\bibinfo {volume} {58}},\
  \bibinfo {pages} {1} (\bibinfo {year} {2010})}\BibitemShut {NoStop}%
\bibitem [{\citenamefont {Chanelière}\ \emph {et~al.}(2010)\citenamefont
  {Chanelière}, \citenamefont {Ruggiero}, \citenamefont {Bonarota},
  \citenamefont {Afzelius},\ and\ \citenamefont {Gouët}}]{Chaneliere2010}%
  \BibitemOpen
  \bibfield  {author} {\bibinfo {author} {\bibfnamefont {T.}~\bibnamefont
  {Chanelière}}, \bibinfo {author} {\bibfnamefont {J.}~\bibnamefont
  {Ruggiero}}, \bibinfo {author} {\bibfnamefont {M.}~\bibnamefont {Bonarota}},
  \bibinfo {author} {\bibfnamefont {M.}~\bibnamefont {Afzelius}}, \ and\
  \bibinfo {author} {\bibfnamefont {J.-L.~L.}\ \bibnamefont {Gouët}},\ }
{\bibfield  {journal}
  {\bibinfo  {journal} {New Journal of Physics}\ }\textbf {\bibinfo {volume}
  {12}},\ \bibinfo {pages} {023025} (\bibinfo {year} {2010})}\BibitemShut
  {NoStop}%
\bibitem [{\citenamefont {Saglamyurek}\ \emph {et~al.}(2011)\citenamefont
  {Saglamyurek}, \citenamefont {Sinclair}, \citenamefont {Jin}, \citenamefont
  {Slater}, \citenamefont {Oblak}, \citenamefont {Bussi{\`e}res}, \citenamefont
  {George}, \citenamefont {Ricken}, \citenamefont {Sohler},\ and\ \citenamefont
  {Tittel}}]{Saglamyurek2011}%
  \BibitemOpen
  \bibfield  {author} {\bibinfo {author} {\bibfnamefont {E.}~\bibnamefont
  {Saglamyurek}}, \bibinfo {author} {\bibfnamefont {N.}~\bibnamefont
  {Sinclair}}, \bibinfo {author} {\bibfnamefont {J.}~\bibnamefont {Jin}},
  \bibinfo {author} {\bibfnamefont {J.}~\bibnamefont {Slater}}, \bibinfo
  {author} {\bibfnamefont {D.}~\bibnamefont {Oblak}}, \bibinfo {author}
  {\bibfnamefont {F.}~\bibnamefont {Bussi{\`e}res}}, \bibinfo {author}
  {\bibfnamefont {M.}~\bibnamefont {George}}, \bibinfo {author} {\bibfnamefont
  {R.}~\bibnamefont {Ricken}}, \bibinfo {author} {\bibfnamefont
  {W.}~\bibnamefont {Sohler}}, \ and\ \bibinfo {author} {\bibfnamefont
  {W.}~\bibnamefont {Tittel}},\ }\href@noop {} {\bibfield  {journal} {\bibinfo
  {journal} {Nature}\ }\textbf {\bibinfo {volume} {469}},\ \bibinfo {pages}
  {512} (\bibinfo {year} {2011})}\BibitemShut {NoStop}%
\bibitem [{\citenamefont {Louisell}\ \emph {et~al.}(1961)\citenamefont
  {Louisell}, \citenamefont {Yariv},\ and\ \citenamefont
  {Siegman}}]{Louisell1961}%
  \BibitemOpen
  \bibfield  {author} {\bibinfo {author} {\bibfnamefont {W.~H.}\ \bibnamefont
  {Louisell}}, \bibinfo {author} {\bibfnamefont {A.}~\bibnamefont {Yariv}}, \
  and\ \bibinfo {author} {\bibfnamefont {A.~E.}\ \bibnamefont {Siegman}},\
  }
{\bibfield  {journal} {\bibinfo
  {journal} {Phys. Rev.}\ }\textbf {\bibinfo {volume} {124}},\ \bibinfo {pages}
  {1646} (\bibinfo {year} {1961})}\BibitemShut {NoStop}%
\bibitem [{\citenamefont {Hong}\ and\ \citenamefont {Mandel}(1985)}]{Hong1985}%
  \BibitemOpen
  \bibfield  {author} {\bibinfo {author} {\bibfnamefont {C.~K.}\ \bibnamefont
  {Hong}}\ and\ \bibinfo {author} {\bibfnamefont {L.}~\bibnamefont {Mandel}},\
  }
{\bibfield  {journal} {\bibinfo
  {journal} {Phys. Rev. A}\ }\textbf {\bibinfo {volume} {31}},\ \bibinfo
  {pages} {2409} (\bibinfo {year} {1985})}\BibitemShut {NoStop}%
\bibitem [{\citenamefont {Ghosh}\ and\ \citenamefont
  {Mandel}(1987)}]{Ghosh1987}%
  \BibitemOpen
  \bibfield  {author} {\bibinfo {author} {\bibfnamefont {R.}~\bibnamefont
  {Ghosh}}\ and\ \bibinfo {author} {\bibfnamefont {L.}~\bibnamefont {Mandel}},\
  }
{\bibfield  {journal} {\bibinfo
   {journal} {Phys. Rev. Lett.}\ }\textbf {\bibinfo {volume} {59}},\ \bibinfo
  {pages} {1903} (\bibinfo {year} {1987})}\BibitemShut {NoStop}%
\bibitem [{\citenamefont {Mandel}(1999)}]{Mandel1999}%
  \BibitemOpen
  \bibfield  {author} {\bibinfo {author} {\bibfnamefont {L.}~\bibnamefont
  {Mandel}},\ }
{\bibfield
  {journal} {\bibinfo  {journal} {Rev. Mod. Phys.}\ }\textbf {\bibinfo {volume}
  {71}},\ \bibinfo {pages} {S274} (\bibinfo {year} {1999})}\BibitemShut
  {NoStop}%
\bibitem [{\citenamefont {Rubin}\ \emph {et~al.}(1994)\citenamefont {Rubin},
  \citenamefont {Klyshko}, \citenamefont {Shih},\ and\ \citenamefont
  {Sergienko}}]{Rubin1994}%
  \BibitemOpen
  \bibfield  {author} {\bibinfo {author} {\bibfnamefont {M.~H.}\ \bibnamefont
  {Rubin}}, \bibinfo {author} {\bibfnamefont {D.~N.}\ \bibnamefont {Klyshko}},
  \bibinfo {author} {\bibfnamefont {Y.~H.}\ \bibnamefont {Shih}}, \ and\
  \bibinfo {author} {\bibfnamefont {A.~V.}\ \bibnamefont {Sergienko}},\ }
{\bibfield  {journal} {\bibinfo
  {journal} {Phys. Rev. A}\ }\textbf {\bibinfo {volume} {50}},\ \bibinfo
  {pages} {5122} (\bibinfo {year} {1994})}\BibitemShut {NoStop}%
\bibitem [{\citenamefont {Keller}\ and\ \citenamefont
  {Rubin}(1997)}]{Keller1997}%
  \BibitemOpen
  \bibfield  {author} {\bibinfo {author} {\bibfnamefont {T.~E.}\ \bibnamefont
  {Keller}}\ and\ \bibinfo {author} {\bibfnamefont {M.~H.}\ \bibnamefont
  {Rubin}},\ }
{\bibfield  {journal}
  {\bibinfo  {journal} {Phys. Rev. A}\ }\textbf {\bibinfo {volume} {56}},\
  \bibinfo {pages} {1534} (\bibinfo {year} {1997})}\BibitemShut {NoStop}%
\bibitem [{\citenamefont {Pittman}\ \emph {et~al.}(1996)\citenamefont
  {Pittman}, \citenamefont {Strekalov}, \citenamefont {Klyshko}, \citenamefont
  {Rubin}, \citenamefont {Sergienko},\ and\ \citenamefont
  {Shih}}]{Pittman1996}%
  \BibitemOpen
  \bibfield  {author} {\bibinfo {author} {\bibfnamefont {T.~B.}\ \bibnamefont
  {Pittman}}, \bibinfo {author} {\bibfnamefont {D.~V.}\ \bibnamefont
  {Strekalov}}, \bibinfo {author} {\bibfnamefont {D.~N.}\ \bibnamefont
  {Klyshko}}, \bibinfo {author} {\bibfnamefont {M.~H.}\ \bibnamefont {Rubin}},
  \bibinfo {author} {\bibfnamefont {A.~V.}\ \bibnamefont {Sergienko}}, \ and\
  \bibinfo {author} {\bibfnamefont {Y.~H.}\ \bibnamefont {Shih}},\ }
{\bibfield  {journal} {\bibinfo
  {journal} {Phys. Rev. A}\ }\textbf {\bibinfo {volume} {53}},\ \bibinfo
  {pages} {2804} (\bibinfo {year} {1996})}\BibitemShut {NoStop}%
\bibitem [{\citenamefont {Joobeur}\ \emph {et~al.}(1994)\citenamefont
  {Joobeur}, \citenamefont {Saleh},\ and\ \citenamefont {Teich}}]{Joobeur1994}%
  \BibitemOpen
  \bibfield  {author} {\bibinfo {author} {\bibfnamefont {A.}~\bibnamefont
  {Joobeur}}, \bibinfo {author} {\bibfnamefont {B.~E.~A.}\ \bibnamefont
  {Saleh}}, \ and\ \bibinfo {author} {\bibfnamefont {M.~C.}\ \bibnamefont
  {Teich}},\ }
{\bibfield  {journal}
  {\bibinfo  {journal} {Phys. Rev. A}\ }\textbf {\bibinfo {volume} {50}},\
  \bibinfo {pages} {3349} (\bibinfo {year} {1994})}\BibitemShut {NoStop}%
\bibitem [{\citenamefont {Kurtsiefer}\ \emph {et~al.}(2001)\citenamefont
  {Kurtsiefer}, \citenamefont {Oberparleiter},\ and\ \citenamefont
  {Weinfurter}}]{Kurtsiefer2001}%
  \BibitemOpen
  \bibfield  {author} {\bibinfo {author} {\bibfnamefont {C.}~\bibnamefont
  {Kurtsiefer}}, \bibinfo {author} {\bibfnamefont {M.}~\bibnamefont
  {Oberparleiter}}, \ and\ \bibinfo {author} {\bibfnamefont {H.}~\bibnamefont
  {Weinfurter}},\ }
{\bibfield
  {journal} {\bibinfo  {journal} {Phys. Rev. A}\ }\textbf {\bibinfo {volume}
  {64}},\ \bibinfo {pages} {023802} (\bibinfo {year} {2001})}\BibitemShut
  {NoStop}%
\bibitem [{\citenamefont {Bovino}\ \emph {et~al.}(2003)\citenamefont {Bovino},
  \citenamefont {Varisco}, \citenamefont {Colla}, \citenamefont {Castagnoli},
  \citenamefont {Giuseppe},\ and\ \citenamefont {Sergienko}}]{Bovino2003}%
  \BibitemOpen
  \bibfield  {author} {\bibinfo {author} {\bibfnamefont {F.~A.}\ \bibnamefont
  {Bovino}}, \bibinfo {author} {\bibfnamefont {P.}~\bibnamefont {Varisco}},
  \bibinfo {author} {\bibfnamefont {A.~M.}\ \bibnamefont {Colla}}, \bibinfo
  {author} {\bibfnamefont {G.}~\bibnamefont {Castagnoli}}, \bibinfo {author}
  {\bibfnamefont {G.~D.}\ \bibnamefont {Giuseppe}}, \ and\ \bibinfo {author}
  {\bibfnamefont {A.~V.}\ \bibnamefont {Sergienko}},\ }
{\bibfield  {journal} 
{\bibinfo  {journal}
  {Optics Communications}\ }\textbf {\bibinfo {volume} {227}},\ \bibinfo
  {pages} {343 } (\bibinfo {year} {2003})}\BibitemShut {NoStop}%
\bibitem [{\citenamefont {Castelletto}\ \emph {et~al.}(2004)\citenamefont
  {Castelletto}, \citenamefont {Degiovanni}, \citenamefont {Migdall},\ and\
  \citenamefont {Ware}}]{Castelletto2004}%
  \BibitemOpen
  \bibfield  {author} {\bibinfo {author} {\bibfnamefont {S.}~\bibnamefont
  {Castelletto}}, \bibinfo {author} {\bibfnamefont {I.~P.}\ \bibnamefont
  {Degiovanni}}, \bibinfo {author} {\bibfnamefont {A.}~\bibnamefont {Migdall}},
  \ and\ \bibinfo {author} {\bibfnamefont {M.}~\bibnamefont {Ware}},\ }
{\bibfield  {journal} {\bibinfo
   {journal} {New Journal of Physics}\ }\textbf {\bibinfo {volume} {6}},\
  \bibinfo {pages} {87} (\bibinfo {year} {2004})}\BibitemShut {NoStop}%
\bibitem [{\citenamefont {Castelletto}\ \emph {et~al.}(2005)\citenamefont
  {Castelletto}, \citenamefont {Degiovanni}, \citenamefont {Furno},
  \citenamefont {Schettini}, \citenamefont {Migdall},\ and\ \citenamefont
  {Ware}}]{Castelletto2005}%
  \BibitemOpen
  \bibfield  {author} {\bibinfo {author} {\bibfnamefont {S.}~\bibnamefont
  {Castelletto}}, \bibinfo {author} {\bibfnamefont {I.}~\bibnamefont
  {Degiovanni}}, \bibinfo {author} {\bibfnamefont {G.}~\bibnamefont {Furno}},
  \bibinfo {author} {\bibfnamefont {V.}~\bibnamefont {Schettini}}, \bibinfo
  {author} {\bibfnamefont {A.}~\bibnamefont {Migdall}}, \ and\ \bibinfo
  {author} {\bibfnamefont {M.}~\bibnamefont {Ware}},\ }\href@noop {} {\bibfield
   {journal} {\bibinfo  {journal} {Instrumentation and Measurement, IEEE
  Transactions on}\ }\textbf {\bibinfo {volume} {54}},\ \bibinfo {pages} {890}
  (\bibinfo {year} {2005})}\BibitemShut {NoStop}%
\bibitem [{\citenamefont {Ljunggren}\ and\ \citenamefont
  {Tengner}(2005)}]{Ljunggren2005}%
  \BibitemOpen
  \bibfield  {author} {\bibinfo {author} {\bibfnamefont {D.}~\bibnamefont
  {Ljunggren}}\ and\ \bibinfo {author} {\bibfnamefont {M.}~\bibnamefont
  {Tengner}},\ }\href@noop {http://link.aps.org/abstract/PRA/v72/e062301} {\bibfield
   {journal} {\bibinfo  {journal} {Phys. Rev. A}\ }\textbf {\bibinfo {volume}
  {72}},\ \bibinfo {eid} {062301} (\bibinfo {year} {2005})}\BibitemShut
  {NoStop}%
\bibitem [{\citenamefont {Ling}\ \emph {et~al.}(2009)\citenamefont {Ling},
  \citenamefont {Chen}, \citenamefont {Fan},\ and\ \citenamefont
  {Migdall}}]{Ling2009}%
  \BibitemOpen
  \bibfield  {author} {\bibinfo {author} {\bibfnamefont {A.}~\bibnamefont
  {Ling}}, \bibinfo {author} {\bibfnamefont {J.}~\bibnamefont {Chen}}, \bibinfo
  {author} {\bibfnamefont {J.}~\bibnamefont {Fan}}, \ and\ \bibinfo {author}
  {\bibfnamefont {A.}~\bibnamefont {Migdall}},\ }\href@noop
  {http://www.opticsexpress.org/abstract.cfm?URI=oe-17-23-21302} {\bibfield
  {journal} {\bibinfo  {journal} {Opt. Express}\ }\textbf {\bibinfo {volume}
  {17}},\ \bibinfo {pages} {21302} (\bibinfo {year} {2009})}\BibitemShut
  {NoStop}%
\bibitem [{\citenamefont {Mitchell}(2009)}]{Mitchell2009}%
  \BibitemOpen
  \bibfield  {author} {\bibinfo {author} {\bibfnamefont {M.~W.}\ \bibnamefont
  {Mitchell}},\ }\href@noop {\doibase 10.1103/PhysRevA.79.043835} {\bibfield
  {journal} {\bibinfo  {journal} {Phys. Rev. A}\ }\textbf {\bibinfo {volume}
  {79}},\ \bibinfo {pages} {043835} (\bibinfo {year} {2009})}\BibitemShut
  {NoStop}%
\bibitem [{\citenamefont {Bennink}(2010)}]{Bennink2010}%
  \BibitemOpen
  \bibfield  {author} {\bibinfo {author} {\bibfnamefont {R.~S.}\ \bibnamefont
  {Bennink}},\ }\href@noop {\doibase 10.1103/PhysRevA.81.053805} {\bibfield
  {journal} {\bibinfo  {journal} {Phys. Rev. A}\ }\textbf {\bibinfo {volume}
  {81}},\ \bibinfo {pages} {053805} (\bibinfo {year} {2010})}\BibitemShut
  {NoStop}%
\bibitem [{\citenamefont {Boyd}(2008)}]{Boyd2008}%
  \BibitemOpen
  \bibfield  {author} {\bibinfo {author} {\bibfnamefont {R.}~\bibnamefont
  {Boyd}},\ }\href@noop {} {\emph {\bibinfo {title} {Nonlinear optics}}}\
  (\bibinfo  {publisher} {Academic Press},\ \bibinfo {year} {2008})\BibitemShut
  {NoStop}%
\bibitem [{\citenamefont {Yariv}(1989)}]{Yariv1989}%
  \BibitemOpen
  \bibfield  {author} {\bibinfo {author} {\bibfnamefont {A.}~\bibnamefont
  {Yariv}},\ }\href@noop {http://books.google.fr/books?id=UTWg1VIkNuMC} {\emph
  {\bibinfo {title} {Quantum electronics}}}\ (\bibinfo  {publisher} {Wiley},\
  \bibinfo {year} {1989})\BibitemShut {NoStop}%
\bibitem [{\citenamefont {Fiorentino}\ \emph {et~al.}(2004)\citenamefont
  {Fiorentino}, \citenamefont {Messin}, \citenamefont {Kuklewicz},
  \citenamefont {Wong},\ and\ \citenamefont {Shapiro}}]{Fiorentino2004}%
  \BibitemOpen
  \bibfield  {author} {\bibinfo {author} {\bibfnamefont {M.}~\bibnamefont
  {Fiorentino}}, \bibinfo {author} {\bibfnamefont {G.}~\bibnamefont {Messin}},
  \bibinfo {author} {\bibfnamefont {C.~E.}\ \bibnamefont {Kuklewicz}}, \bibinfo
  {author} {\bibfnamefont {F.~N.~C.}\ \bibnamefont {Wong}}, \ and\ \bibinfo
  {author} {\bibfnamefont {J.~H.}\ \bibnamefont {Shapiro}},\ }\href@noop {\doibase
  10.1103/PhysRevA.69.041801} {\bibfield  {journal} {\bibinfo  {journal} {Phys.
  Rev. A}\ }\textbf {\bibinfo {volume} {69}},\ \bibinfo {eid} {041801}
  (\bibinfo {year} {2004})}\BibitemShut {NoStop}%
\bibitem [{\citenamefont {Hentschel}\ \emph {et~al.}(2009)\citenamefont
  {Hentschel}, \citenamefont {H\"{u}bel}, \citenamefont {Poppe},\ and\
  \citenamefont {Zeilinger}}]{Hentschel2009}%
  \BibitemOpen
  \bibfield  {author} {\bibinfo {author} {\bibfnamefont {M.}~\bibnamefont
  {Hentschel}}, \bibinfo {author} {\bibfnamefont {H.}~\bibnamefont
  {H\"{u}bel}}, \bibinfo {author} {\bibfnamefont {A.}~\bibnamefont {Poppe}}, \
  and\ \bibinfo {author} {\bibfnamefont {A.}~\bibnamefont {Zeilinger}},\ }\href@noop
  {http://www.opticsexpress.org/abstract.cfm?URI=oe-17-25-23153} {\bibfield
  {journal} {\bibinfo  {journal} {Opt. Express}\ }\textbf {\bibinfo {volume}
  {17}},\ \bibinfo {pages} {23153} (\bibinfo {year} {2009})}\BibitemShut
  {NoStop}%
\bibitem [{\citenamefont {Garrison}\ and\ \citenamefont
  {Chiao}(2008)}]{Garrison2008}%
  \BibitemOpen
  \bibfield  {author} {\bibinfo {author} {\bibfnamefont {J.}~\bibnamefont
  {Garrison}}\ and\ \bibinfo {author} {\bibfnamefont {R.}~\bibnamefont
  {Chiao}},\ }\href@noop {} {\emph {\bibinfo {title} {Quantum optics}}}\
  (\bibinfo  {publisher} {Oxford University Press},\ \bibinfo {year}
  {2008})\BibitemShut {NoStop}%
\bibitem [{\citenamefont {Shapiro}(2009)}]{Shapiro2009}%
  \BibitemOpen
  \bibfield  {author} {\bibinfo {author} {\bibfnamefont {J.}~\bibnamefont
  {Shapiro}},\ }\href@noop {\doibase 10.1109/JSTQE.2009.2024959} {\bibfield
  {journal} {\bibinfo  {journal} {Selected Topics in Quantum Electronics, IEEE
  Journal of}\ }\textbf {\bibinfo {volume} {15}},\ \bibinfo {pages} {1547 }
  (\bibinfo {year} {2009})}\BibitemShut {NoStop}%
\bibitem [{Note1()}]{Note1}%
  \BibitemOpen
  \bibinfo {note} {The natural phase matching bandwidth has no influence on the
  results that follow if it is much larger than the pump
  linewidth.}\BibitemShut {Stop}%
\bibitem [{\citenamefont {Berntsen}\ \emph {et~al.}(1991)\citenamefont
  {Berntsen}, \citenamefont {Espelid},\ and\ \citenamefont
  {Genz}}]{Berntsen1991}%
  \BibitemOpen
  \bibfield  {author} {\bibinfo {author} {\bibfnamefont {J.}~\bibnamefont
  {Berntsen}}, \bibinfo {author} {\bibfnamefont {T.~O.}\ \bibnamefont
  {Espelid}}, \ and\ \bibinfo {author} {\bibfnamefont {A.}~\bibnamefont
  {Genz}},\ }\href@noop {\doibase http://doi.acm.org/10.1145/210232.210233}
  {\bibfield  {journal} {\bibinfo  {journal} {ACM Trans. Math. Softw.}\
  }\textbf {\bibinfo {volume} {17}},\ \bibinfo {pages} {437} (\bibinfo {year}
  {1991})}\BibitemShut {NoStop}%
\bibitem [{\citenamefont {Boyd}\ and\ \citenamefont
  {Kleinman}(1968)}]{Boyd1968}%
  \BibitemOpen
  \bibfield  {author} {\bibinfo {author} {\bibfnamefont {G.~D.}\ \bibnamefont
  {Boyd}}\ and\ \bibinfo {author} {\bibfnamefont {D.~A.}\ \bibnamefont
  {Kleinman}},\ }\href@noop {http://link.aip.org/link/?JAP/39/3597/1} {\bibfield
  {journal} {\bibinfo  {journal} {J. Appl. Phys.}\ }\textbf {\bibinfo {volume}
  {39}},\ \bibinfo {pages} {3597} (\bibinfo {year} {1968})}\BibitemShut
  {NoStop}%
\bibitem [{\citenamefont {Smirr}\ \emph
  {et~al.}(2011{\natexlab{b}})\citenamefont {Smirr}, \citenamefont {Guilbaud},
  \citenamefont {Ghalbouni}, \citenamefont {Frey}, \citenamefont {Diamanti},
  \citenamefont {All\'{e}aume},\ and\ \citenamefont {Zaquine}}]{Smirr2010b}%
  \BibitemOpen
  \bibfield  {author} {\bibinfo {author} {\bibfnamefont {J.~L.}\ \bibnamefont
  {Smirr}}, \bibinfo {author} {\bibfnamefont {S.}~\bibnamefont {Guilbaud}},
  \bibinfo {author} {\bibfnamefont {J.}~\bibnamefont {Ghalbouni}}, \bibinfo
  {author} {\bibfnamefont {R.}~\bibnamefont {Frey}}, \bibinfo {author}
  {\bibfnamefont {E.}~\bibnamefont {Diamanti}}, \bibinfo {author}
  {\bibfnamefont {R.}~\bibnamefont {All\'{e}aume}}, \ and\ \bibinfo {author}
  {\bibfnamefont {I.}~\bibnamefont {Zaquine}},\ }\href@noop {\doibase
  10.1364/OE.19.000616} {\bibfield  {journal} {\bibinfo  {journal} {Opt.
  Express}\ }\textbf {\bibinfo {volume} {19}},\ \bibinfo {pages} {616}
  (\bibinfo {year} {2011}{\natexlab{b}})}\BibitemShut {NoStop}%
\bibitem [{\citenamefont {Ling}\ \emph {et~al.}(2008)\citenamefont {Ling},
  \citenamefont {Lamas-Linares},\ and\ \citenamefont {Kurtsiefer}}]{Ling2008}%
  \BibitemOpen
  \bibfield  {author} {\bibinfo {author} {\bibfnamefont {A.}~\bibnamefont
  {Ling}}, \bibinfo {author} {\bibfnamefont {A.}~\bibnamefont {Lamas-Linares}},
  \ and\ \bibinfo {author} {\bibfnamefont {C.}~\bibnamefont {Kurtsiefer}},\
  }\href@noop {\doibase 10.1103/PhysRevA.77.043834} {\bibfield  {journal} {\bibinfo
  {journal} {Phys. Rev. A}\ }\textbf {\bibinfo {volume} {77}},\ \bibinfo
  {pages} {043834} (\bibinfo {year} {2008})}\BibitemShut {NoStop}%
\bibitem [{\citenamefont {Mandel}(1966)}]{Mandel1966}%
  \BibitemOpen
  \bibfield  {author} {\bibinfo {author} {\bibfnamefont {L.}~\bibnamefont
  {Mandel}},\ }\href@noop {\doibase 10.1103/PhysRev.144.1071} {\bibfield  {journal}
  {\bibinfo  {journal} {Phys. Rev.}\ }\textbf {\bibinfo {volume} {144}},\
  \bibinfo {pages} {1071} (\bibinfo {year} {1966})}\BibitemShut {NoStop}%
\bibitem [{\citenamefont {Ghatak}\ and\ \citenamefont
  {Thyagarajan}(1998)}]{Ghatak1998}%
  \BibitemOpen
  \bibfield  {author} {\bibinfo {author} {\bibfnamefont {A.}~\bibnamefont
  {Ghatak}}\ and\ \bibinfo {author} {\bibfnamefont {K.}~\bibnamefont
  {Thyagarajan}},\ }\href@noop {} {\emph {\bibinfo {title} {Introduction to
  fiber optics}}}\ (\bibinfo  {publisher} {Cambridge University Press},\
  \bibinfo {year} {1998})\BibitemShut {NoStop}%
\end{thebibliography}
\end{document}